\documentstyle[preprint,eqsecnum,prd,aps]{revtex}
\begin{document} 
\draft
%%%%%End of Preamble
%%%%Start of Text%%%%%%%%%%%%%%%%%%%%%%%%%%%%%%%%%%%%%%%%%%%%%%%%%%%%%%%
\preprint{
\vbox{\openup -3\jot
\halign{&##\hfil\cr
	& ANL-HEP-PR-95-59 \cr
	& (REVISED)\cr	
	& September 1996 \cr}}
}
\title{
A Lattice Formulation of Chiral Gauge Theories \\
}
\author{Geoffrey T. Bodwin}
\address{
High Energy Physics Division, Argonne National Laboratory,
Argonne, IL 60439
}
\maketitle
\begin{abstract} 
We present a method for formulating gauge theories of chiral fermions in
lattice field theory.  The method makes use of a Wilson mass to remove
doublers.  Gauge invariance is then restored by modifying the theory in
two ways: the magnitude of the fermion determinant is replaced with the
square root of the determinant for a fermion with vector-like couplings
to the gauge field; a double limit is taken, in which the lattice
spacing associated with the fermion field is sent to zero before the
lattice spacing associated with the gauge field. The method applies only
to theories whose fermions are in an anomaly-free representation of the
gauge group. We also present a related technique for computing matrix
elements of operators involving fermion fields.  Although the analyses
of these methods are couched in weak-coupling perturbation theory, it is
argued that the computational prescriptions are gauge invariant in the
presence of a nonperturbative gauge-field configuration. 
\end{abstract}

\narrowtext
\section{INTRODUCTION}
\label{sec:introduction}

The interaction of chiral spin-1/2 particles with gauge fields is a feature 
of many field-theoretic models, including the standard electroweak model.  
The implementation of chiral gauge theories in lattice field theory is, 
of course, a prerequisite to the numerical simulation of such theories; 
but it is also of importance in establishing that chiral theories can be
defined outside of the domain of perturbation theory.

In recent years, a number of proposals for constructing lattice versions 
of chiral gauge theories have been put forward.  A review of the present
status of many of these lattice chiral-fermion proposals has been given 
by Shamir \cite{shamir}.  Some proposals 
\cite{slavnov,zenkin,bietenholz-weise} have not yet been studied
extensively.  Others, such as the Eichten-Preskill model 
\cite{eichten-preskill}, the Smit-Swift model \cite {smit-swift}, and the 
staggered-fermion model \cite{smit,staggered-fermion},
apparently fail to yield a chiral fermion spectrum because of the
coupling of gauge degrees of freedom to the fermion
\cite{eichten-preskill-n,smit-swift-n}. The domain-wall proposal of
Kaplan \cite{domain-wall} and the related overlap formula of Narayanan
and Neuberger \cite{narayanan-neuberger} have received a good deal of
study, with encouraging results 
\cite{narayanan-neuberger,domain-wall-p,overlap-p}.  On the
other hand, it has been suggested that both of these methods might fail 
along the lines of the failure of the Smit-Swift model because gauge 
degrees of freedom couple to the fermion at the boundaries of the
regions of nonzero gauge field \cite{domain-wall-n,overlap-n}.
Given the unsettled status of the proposals that are currently viable, 
it would seem to be worthwhile to consider alternative methods for 
formulating chiral gauge theories.

In this paper, we present a new method for constructing lattice versions
of chiral gauge theories.  Our approach makes use of a Wilson mass
\cite{wilson} to remove fermion species doublers.  The Wilson mass
breaks the chiral gauge symmetry.  However, we argue that the violations 
of chiral symmetry that survive in the continuum limit are associated 
with ultraviolet- (UV-) divergent amplitudes and that the chiral
symmetry can be partially restored through the addition of local
renormalization counterterms to the action \cite{lat90,dpf91,lat92}. 
The philosophy of using local counterterms to restore the chiral
symmetry has also been suggested by the Rome group \cite{rome}. (Local
counterterms are required to restore the chiral symmetry in the proposal
of the Zaragoza group \cite{zaragoza}, as well.) However, unlike the
approach of the Rome group, our method does not entail the tuning of
counterterm coefficients. Instead, we implement the counterterms by 
modifying the lattice definitions of the fermion determinant and 
operator matrix elements.

The first modification is to replace the magnitude of the fermion
determinant with the square root of the determinant of a fermion with
{\it vector-like} couplings to the gauge field
\cite{lat90,dpf91,lat92,alvarez-gaume-pietra,gockeler-schierholz}. (A
related modification of matrix elements of operators involving fermion
fields is also introduced.) This redefinition of the determinant
implements the renormalization counterterms that are associated with UV
divergences in a single fermion loop. After this modification, the
fermion determinant is gauge invariant in the presence of a background
gauge field, except for contributions from the Adler-Bardeen-Jackiw
(ABJ) anomaly \cite{effective-actions}. These violations of chiral
symmetry cancel, as usual, when one considers a theory containing a
suitable complement of physical fermions. 

The presence of dynamical gauge fields leads to additional ultraviolet
divergences and potentially requires the introduction of many new
counterterms to restore the chiral gauge symmetry. We deal with this 
difficulty by introducing separate lattice cutoffs for the fermion
fields and gauge fields 
\cite{smit,gockeler-schierholz,anlreport,thooft,hsu,kronfeld}.
In the double limit in which the fermion cutoff is removed before the
gauge-field cutoff, the violations of chiral symmetry vanish with at
least one power of the ratio of cutoffs. The use of this double limit in 
conjunction with the modification of the magnitude of the fermion 
determinant has been emphasized previously in 
Refs.~\cite{gockeler-schierholz,anlreport}

Most of the analysis in this paper is couched in weak-coupling
coupling perturbation theory.  However, we are able to show, by
exploiting the finite radius of convergence the perturbation expansion
of the fermion determinant, that our method is also valid in the
presence of nonperturbative gauge-field configurations. 

The remainder of this paper is organized as follows.  In
Section~\ref{sec:doubling} we discuss, in general terms, fermion
doubling, its elimination through the use of a Wilson mass, and the
breaking and restoration of chiral symmetry.  In Section~\ref{sec:model}
we introduce a lattice implementation of a theory of left-handed
fermions coupled to a non-Abelian gauge field. Although our specific
analyses in subsequent sections of the paper refer to this model, our
methods generalize immediately to models that contain right-handed as
well as left-handed fermion fields and to models that contain scalar
particles.  In Section~\ref{sec:gauge} we discuss the nature of the
violations of gauge invariance that arise from the introduction of a
Wilson mass. Section~\ref{sec:background-gauge} contains an analysis of
the chiral-symmetry properties of the fermion determinant in the
presence of a background gauge field.  This analysis allows us to derive
a modification of the determinant that restores the chiral symmetry in
the case of an anomaly-free theory. In Section~\ref{sec:dynamical} we
discuss the difficulties that arise from dynamical gauge fields and
present the double-limit procedure for dealing with them. In
Section~\ref{sec:operators} we indicate how the methods used in
computing the fermion determinant can also be applied in computing
matrix elements of operators containing fermion fields. A proof of the
validity of the methods for computing the fermion determinant and
operator matrix elements in the presence of nonperturbative gauge fields
is sketched in Section~\ref{sec:nonpert}. Finally, in
Section~\ref{sec:summary}, we summarize our results and discuss various
options for implementing our chiral-fermion method. 

While this paper was in preparation, a paper by Hern\'andez and Sundrum
\cite{hernandez-sundrum} on the same subject appeared. The methods that
these authors propose for computing the fermion determinant (but not the
matrix elements of fermion operators) are essentially identical to the
ones proposed in the present paper.  Many of the conclusions drawn in
the present paper and in Ref.~\cite{hernandez-sundrum} are the same; one
exception is noted at the end of Section~\ref{sec:counting-a-f}. 
However, the details of the proofs in the two papers are, in general,
quite different. 

\section{DOUBLING, WILSON MASSES, CHIRAL SYMMETRY: GENERAL 
CONSIDERATIONS}
\label{sec:doubling}

It is well known that the most straightforward transcription of the 
Dirac operator to the lattice is problematic because of the phenomenon
of fermion doubling: for each left- or right-handed particle in the
continuum theory, there are $2^{d-1}$ left-handed and $2^{d-1}$
right-handed particles in the lattice theory, where $d$ is the
dimensionality of space-time \cite{karsten-smit}.  

For the case of QCD, Wilson \cite{wilson} suggested that one could
remove the doublers by giving them a mass that goes to infinity as the
lattice spacing $a$ goes to zero.  Of course, the introduction of a mass
explicitly breaks the chiral symmetry.  However, this is not expected to
present a serious problem in QCD, since the gauge symmetry remains
intact. Consequently, the renormalization program is unaffected and one
should recover the continuum theory as the lattice regulator is removed
($a\rightarrow 0$). 

In the case of a chiral gauge theory, the introduction of a Wilson mass
has more serious consequences.  For such a theory, the Wilson mass and,
hence, the UV regulator break the gauge symmetry, thereby jeopardizing
the renormalization program and the decoupling of unphysical degrees of
freedom. A failure of the gauge degrees of freedom to decouple may lead
to an alteration of the low-energy spectrum of the theory. For example,
under such circumstances, when one integrates over the gauge degrees of
freedom, a chiral gauge theory can become a vector-like gauge theory
\cite{smit-swift-n}. 

In a chiral theory, one cannot completely avoid such a breaking of the
gauge symmetry. There are several no-go theorems which state, under a
variety of assumptions, that any gauge theory that does not exhibit
fermion doubling must violate chiral symmetry \cite{karsten-smit,no-go}.
One can argue this very generally on the basis of the properties of the
ABJ anomaly.  If a lattice theory preserves a chiral symmetry, then the
corresponding chiral current is conserved.  In particular, the triangle
anomaly is zero and remains zero in the continuum limit.  But, according
to the proof of Adler and Bardeen \cite{adler-bardeen}, there is no
Lorentz-covariant Bose-symmetric counterterm that removes the anomaly in
the triple-chiral-current Green's function for a theory containing a
single fermion species. That is, there is no UV regulator under which
the anomaly vanishes as the regulator is removed. Hence, a lattice
regulator that preserves the chiral symmetry must cancel the anomaly
through the presence of multiple fermion species, {\it i.e.} doubling.
Note that this argument leaves open the possibility that one might
eliminate the doubling in a way such that the violations of chiral
symmetry arise {\it solely} from the ABJ anomaly.  Such a result is our
goal. 

In employing continuum perturbative UV regulators, such as dimensional
regularization, one deals with violations of a chiral gauge symmetry by
adding counterterms order by order in perturbation theory so as to
restore the chiral symmetry in selected Green's functions.  The
remaining violations of the chiral symmetry arise from the ABJ anomaly
and cancel when one introduces an appropriate complement of physical
fermion species. Such an order-by-order approach is, of course,
incompatible with a nonperturbative regularization of the theory.
However, one might still hope to effect a restoration of the chiral
symmetry by introducing local counterterms with appropriate
coefficients. 

A heuristic argument in support of this idea is the following.  Suppose
that we have introduced a Wilson mass term.  Then, the lattice spectrum
for the non-interacting theory is identical to the continuum spectrum in
the limit $a\rightarrow 0$. Suppose also that we have fixed to a
renormalizable gauge.  Then, the magnitude of the gauge field is much
less than order $1/a$, unless a source of the gauge field has momentum of
order $1/a$. Consequently, for field momenta much less than $1/a$, the
interacting lattice action approaches the continuum action in the limit
$a\rightarrow 0$. The conclusion is that the lattice, in this case, is
simply a UV regulator.  It follows that the differences between the
lattice regularization and any other UV regularization must reside at
loop momenta on the order of the UV cutoff of the theory.  Hence, the
differences must arise at short distances ($\sim 1/\hbox{cutoff}$); that
is, they have the form of local interactions.  Therefore, we 
conclude that, if there exists a satisfactory UV regularization of a
chiral gauge theory (that is, one that respects the chiral gauge
symmetry), then it must be equivalent to the (Wilson)
lattice-regularized theory, plus local counterterms. Furthermore, if we 
find such a theory, it is unique, up to gauge-invariant counterterms, 
which merely renormalize the coupling constant.

\section{A LATTICE CHIRAL-FERMION MODEL}
\label{sec:model}

Now let us discuss the lattice implementation of a specific model: a 
left-handed fermion coupled to a non-Abelian gauge field.  As we have 
already mentioned, the techniques that we present are easily 
generalizable to models containing right-handed fermions and/or scalar 
particles.  

We assume that the gauge-field part of the (Euclidean) action has the
standard plaquette form 
\begin{equation}
S_G={1\over 2 g^2}\sum_x\sum_{\mu\neq\nu} {\rm Tr}\, U_\mu(x)U_\nu(x+a_\mu)
U_\mu^\dagger(x+a_\nu)U_\nu^\dagger(x)+\hbox{h.c.},
\label{gauge-field-action}
\end{equation}
where, as usual, 
\begin{mathletters}
\label{plaquette}
\begin{equation}
U_\mu(x)\equiv \exp\left[iagA_\mu(x+a_\mu/2)\right],
\end{equation}
\begin{equation}
U_{\mu}^\dagger(x)\equiv \exp\left[-iagA_\mu(x+a_\mu/2)\right]
\end{equation}
\end{mathletters}%
are the lattice link variables, $A_\mu=A_\mu^a T_a$ is the gauge field,
$T_a$ is a gauge-group matrix in the fundamental representation, $g$
is the gauge-field coupling, $a$ is the lattice spacing, and $a_\mu$ is a
unit vector in the $\mu$~direction. Initially, we introduce the fermion
through the ``naive'' lattice action for a Dirac particle: 
\begin{equation}
S_N=a^d\sum_{x,\mu}\overline\psi(x)\gamma_\mu{1\over 
2a}[\psi(x+a_\mu)-\psi(x-a_\mu)],
\label{naive-action}
\end{equation}
where the $\gamma$'s are Euclidean Dirac matrices satisfying 
$\{\gamma_\mu,\gamma_\nu\}=2\delta_{\mu\nu}$.
Note that, in contrast with some formulations of chiral theories, our
approach retains both left- and right-handed components in the fermion
field. The chiral nature of the theory arises from the coupling to gauge
fields, which involves only the left-handed Dirac component: 
\begin{equation}
{S_{NI}=a^d\sum_{x,\mu}\overline\psi(x)\gamma_\mu P_L
{1\over 2a}
\{[U_\mu(x)-1]\psi(x+a_\mu)}
-[U_\mu^\dagger(x-a_\mu)-1]\psi(x-a_\mu)\},
\label{naive-gauge}
\end{equation}
where $P_{R/L}=(1/2)(1\pm \gamma_5)$, $\{\gamma_5,\gamma_\mu\}=0$,
$\gamma_5^2=1$. (In four dimensions,
$\gamma_5=-\gamma_1\gamma_2\gamma_3\gamma_4$.) 
The fermion propagator corresponding to the naive action is 
\begin{equation}
iS_F^N(p)=[(1/a) \sum_\mu i\gamma_\mu \sin(p_\mu a)]^{-1},
\label{naiveprop}\end{equation}
where $p$ is the incoming fermion momentum. The order $g$ and order
$g^2$ gauge-field vertices that arise from the gauging of the naive
action are 
\begin{mathletters}
\begin{equation}
{\cal V}_{\mu,a}^{(1)N}(p,l)=T_a V_{\mu}^{(1)N}(p,l) P_L
=-igT_a\gamma_\mu P_L\cos[(p_\mu+\case 1/2 l_\mu)a],
\end{equation}
\begin{equation}
{\cal V}_{\mu\nu,ab}^{(2)N}(p,l_1,l_2)=T_a T_b 
V_{\mu\nu}^{(2)N}(p,l_1,l_2) P_L
=iag^2T_aT_b\delta_{\mu\nu}\gamma_\mu P_L\sin[(p_\mu
+\case 1/2 l_{1\mu}+\case 1/2 l_{2\mu})a],
\end{equation}
\end{mathletters}%
where the $V^N$'s are the vertices that arise from the gauging of the
naive lattice action for a theory of fermions with vector-like couplings
to an Abelian gauge field.  Here $T_a,T_b,\ldots$ are the gauge-group
matrices, $a,b,\ldots$ are the gauge-field indices, $\mu,\nu\ldots$ are
the polarization indices, and $l_1,l_2,\ldots$ are the incoming momenta,
all of which are associated respectively with the gauge
fields.  The incoming fermion momentum is $p$. The vertices of higher
order in $g$ can be obtained conveniently from the recursion relation 
\begin{eqnarray}
{V}_{\mu_1 \ldots \mu_{n+1}}^{(n+1)}&&(p,l_1,\ldots,l_{n+1})\nonumber\\
&&=-g \delta_{\mu_n \mu_{n+1}} {{V}_{\mu_1
\ldots \mu_n}^{(n)}(p+l_{n+1},l_1,\ldots,l_n)
-{V}_{\mu_1 \ldots \mu_n}^{(n)}(p,l_1,\ldots,l_n)
\over d_{\mu_{n+1}} (l_{n+1})},
\label{recursion}
\end{eqnarray}
where 
\begin{equation}
d_\mu=(2/a)\sin \case 1/2 p_{\mu} a.
\label{lat-mom}
\end{equation}

In addition to the usual pole at $p=0$, the naive propagator
(\ref{naiveprop}) has extra poles when one or more momentum components
are equal to $\pi/a$.  It can be seen that half of the poles have
positive chiral charge and half have negative chiral charge
\cite{karsten-smit}. Thus, this doubling phenomenon leads to gauge-field
couplings to both left- and right-handed species; the theory, at this 
stage, is not chiral. 

We follow the standard approach of eliminating the doublers by including 
a Wilson mass term \cite{wilson} in the action:
\begin{equation}
{S_W=a^d\sum_{x,\mu}
\overline\psi(x)
{1\over 2a}[2\psi(x)-\psi(x+a_\mu)}
-\psi(x-a_\mu)].
\label{wilson-action}
\end{equation}
We can gauge the Wilson term by adding to the action
\begin{equation}
{S_{WI}=a^d\sum_{x,\mu}\overline\psi(x) {1\over 2a}
\{[1-U_\mu(x)]\psi(x+a_\mu)}
+[1-U_\mu^\dagger(x-a_\mu)]\psi(x-a_\mu)\}.
\label{wilson-gauge}
\end{equation}
(As we shall see, it may sometimes be convenient to drop this coupling
of the Wilson term to the gauge field.) 

Now the fermion propagator is 
\begin{equation}
i{S_F^W(p)=\{(1/a)\sum_\mu i\gamma_\mu\sin(p_\mu a)}
+M(p)\}^{-1},
\label{prop-wilson}
\end{equation}
where $M(p)$ is the Fourier transform of the Wilson mass:
\begin{equation}
M(p)=(1/a)\sum_\mu[1-\cos(p_\mu a)].
\label{wilson-mass}
\end{equation}
The additional vertices that arise from the gauging 
of the Wilson term are
\begin{mathletters}
\label{vert-wilson}
\begin{equation}
{\cal V}_{\mu,a}^{(1)W}(p,l)=T_a V_{\mu}^{(1)W}(p,l) 
=-gT_a\sin[(p_\mu+\case 1/2 l_\mu)a],
\end{equation}
\begin{equation}
{\cal V}_{\mu\nu,ab}^{(2)W}(p,l_1,l_2)=T_a T_b
V_{\mu\nu}^{(2)W}(p,l_1,l_2)
=ag^2T_aT_b\delta_{\mu\nu}\gamma_\mu \cos[(p_\mu
+\case 1/2 l_{1\mu}+\case 1/2 l_{2\mu})a],
\end{equation}
\end{mathletters}%
where the higher-order contributions can again be obtained from the
recursion relation (\ref{recursion}).  

We see that the propagator (\ref{prop-wilson}) now has a pole only at
$p=0$.  This would seem to leave us, as desired, with a single Dirac
particle with only left-handed couplings to the gauge field. 
Unfortunately, the Wilson terms $S_W$ and $S_{WI}$, having the Dirac
structures of masses, lead to a nonconservation of the left-handed
vector current by coupling the right-handed component of the Dirac field
back into the theory.  This implies that the chiral gauge invariance of
the theory is broken. 

Such violations of the chiral gauge symmetry cause serious difficulties.
Gauge invariance is an important ingredient in the standard
renormalization program.  Without it, there is an explosion of new
counterterms.  For example, in the absence of current conservation, the
vacuum polarization can generate a quadratically divergent gauge-boson
mass, the light-by-light graph requires counterterms,
Lorentz-noncovariant counterterms can arise on the lattice, and, in
non-Abelian theories, the gauge-boson--fermion coupling can become
different from the triple-gauge-boson coupling. In order to recover a
satisfactory theory of chiral fermions coupled to massless gauge bosons,
one would need to tune all of these counterterms in such a way as to
restore the chiral current conservation.  This is required, for example,
to obtain a massless gauge boson and to guarantee that ghost fields
decouple and that unitarity is preserved. 

On the other hand, we note that the Wilson mass (\ref{wilson-mass}) and
vertices (\ref{vert-wilson}) have the property that they vanish in the
continuum limit $a\rightarrow 0$ for fixed momenta: they are lattice
artifacts.  Consequently, we expect the violations of the gauge symmetry
generated by the Wilson mass to vanish, except when momenta of the order
the lattice cutoff $\pi/a$ are important. That is, we expect that, in
the continuum limit, the violations of the chiral gauge symmetry in
the Green's functions of the theory will persist only in 
UV-divergent Feynman diagrams and subdiagrams. 

\section{GAUGE VARIATIONS}
\label{sec:gauge}

In order to test this expectation, let us examine in more detail the
nature of the violations of the gauge symmetry that result from the
introduction of a Wilson mass.  An infinitesimal transformation of the
gauge field 
\begin{mathletters}
\label{gauge-tx}
\begin{eqnarray}
U_\mu(x)&\rightarrow& U_\mu(x)+i\Lambda(x)U_\mu(x)
-iU_\mu(x)\Lambda(x+a_\mu),
\nonumber\\
U_\mu^\dagger(x)&\rightarrow& 
U_\mu^\dagger(x)+i\Lambda(x+a_\mu)U_\mu^\dagger(x)
-iU_\mu^\dagger(x)\Lambda(x)
\label{u-gauge-tx}
\end{eqnarray}
can be compensated, so as to leave $S_N+S_{NI}$ unchanged, by a 
transformation of the left-handed component of the fermion field:
\begin{eqnarray}
\psi(x)&\rightarrow& [1+iP_L\Lambda(x)]\psi(x),\nonumber\\
\overline\psi(x)&\rightarrow& \overline\psi(x)[1-iP_R\Lambda(x)].
\label{psi-gauge-tx}
\end{eqnarray}
\end{mathletters}%

The Wilson terms, however, are not invariant under the transformation 
(\ref{gauge-tx}).  The gauge transformation results in a change in the 
action:
\begin{eqnarray}
\delta(S_W+S_{WI})&=& a^d\sum_{x,\mu}\overline\psi(x){i\over 2a}
\{2(P_L-P_R)\Lambda(x)\psi(x)\nonumber\\
&&\qquad -[(1-P_R)\Lambda(x)U_\mu(x)-(1-P_L)U_\mu(x)\Lambda(x+a_\mu)]
\psi(x+a_\mu)\nonumber\\
&&\qquad -[(1-P_R)\Lambda(x)U_\mu^\dagger(x-a_\mu)\nonumber\\
&&\qquad\qquad
-(1-P_L)U_\mu^\dagger(x-a_\mu)\Lambda(x-a_\mu)]\psi(x-a_\mu)\}.
\label{gauge-var}
\end{eqnarray}
By Fourier transforming (\ref{gauge-var}), one can arrive at the 
Feynman rules for the vertices corresponding to a gauge variation.  
There is a $\Lambda$-fermion vertex
\begin{mathletters}
\label{var-rules}
\begin{equation}
{\cal M}^{(0)}(p,k)=-iT_a(1-P_R)M(p)+iT_a(1-P_L)M(p+k),
\label{var-mass-vert}
\end{equation}
and there are $\Lambda$-gauge-field-fermion vertices involving 
$n\geq 1$ gauge bosons,
\begin{eqnarray}
{\cal M}_{\mu_1\ldots\mu_n,a_1\ldots a_n}^{(n)}(p,k,l_1,\ldots,l_n)&=&
-iT_a(1-P_R){\cal V}_{\mu_1\ldots\mu_n,a_1\ldots a_n}^{(n)W}
(p,l_1,\ldots,l_n)\nonumber\\
&&\qquad +iT_a(1-P_L){\cal V}_{\mu_1\ldots\mu_n,a_1\ldots a_n}^{(n)W}
(p+k,l_1,\ldots,l_n).
\label{var-vertices}
\end{eqnarray}
\end{mathletters}%
Here, $T_a$ is the gauge-group matrix associated with the gauge
transformation $\Lambda$, $k$ is the incoming momentum associated with
the gauge transformation, $p$ is the incoming fermion momentum, and the
$l_i$ are the incoming gauge-field momenta.  Note that the
$\Lambda$~vertices (\ref{var-rules}) contain factors of $g$ only for
the gauge fields, not for the $\Lambda$~fields. 

If we choose not to gauge the Wilson term, then all of the gauge
variation in the action resides in $S_W$: 
\begin{eqnarray}
\delta(S_W)&=& a^d\sum_{x,\mu}\overline\psi(x){i\over 2a}
\{2(P_L-P_R)\Lambda(x)\psi(x)\nonumber\\
&&\qquad -[-P_R\Lambda(x)+P_L \Lambda(x+a_\mu)]
\psi(x+a_\mu)\nonumber\\
&&\qquad -[-P_R\Lambda(x)
+P_L \Lambda(x-a_\mu)]\psi(x-a_\mu) \}.
\label{gauge-var-wilson}
\end{eqnarray}
In this case, there is a slightly different $\Lambda$-fermion vertex,
\begin{equation}
\tilde{\cal M}^{(0)}(p,k)=iT_aP_R M(p)-iT_aP_L M(p+k),
\label{var-nogauge-mass-vert}
\end{equation}
and there are no $\Lambda$-gauge-field-fermion vertices.

In the analysis to follow, we will frequently make use of the fact that
theories with vector-like couplings to the gauge field exhibit a gauge
invariance, even in the presence of a Wilson mass term. A theory with
vector-like couplings to the gauge field can be obtained by setting
$P_R=P_L=1$ in the action
(\ref{naive-action}), (\ref{naive-gauge}), (\ref{wilson-action}),
(\ref{wilson-gauge}).  Then, if one sets $P_R=P_L=1$ in the gauge
transformation (\ref{gauge-tx}), the gauge variation (\ref{gauge-var})
and the $\Lambda$~vertices (\ref{var-rules}) vanish, as expected.
Note, however, that the gauge symmetry is violated if one drops the
gauging of the Wilson term (\ref{wilson-gauge}) from the action, as can
be seen from examination of (\ref{gauge-var-wilson}) and
(\ref{var-nogauge-mass-vert}). 

There is also a property of the $\Lambda$~vertices that will be crucial 
for our subsequent analysis.  The $\Lambda$~vertices are linear 
combinations of either Wilson masses or Wilson vertices.  Consequently, 
they all vanish in the continuum limit $a \rightarrow 0$ for fixed 
momenta.  Thus, the gauge variations can persist in the limit $a 
\rightarrow 0$ only if momenta of order the lattice cutoff $\pi/a$ are 
important, that is, only in divergent Feynman diagrams.

\section{AMPLITUDES IN A BACKGROUND GAUGE FIELD}
\label{sec:background-gauge}
As a first step in identifying and dealing with the violations of gauge
symmetry in the Green's functions of the chiral theory, let us consider
the case of fermion amplitudes in the presence of background gauge
fields in which the momentum of a gauge-field quantum is limited be much 
less in  magnitude than the lattice momentum cutoff $\pi/a$. 

\subsection{Counting powers of $a$}
\label{sec:divergent}
First let us consider, in the limit $a\rightarrow 0$, the size of the
contribution from a fermion loop containing zero or one gauge-variation
($\Lambda$) vertices and any number of background gauge-field vertices.
We will analyze, in turn, the region of integration in which the
magnitude of the loop momentum is much smaller than $\pi/a$ and the
region of integration in which the magnitude of the loop momentum is of
order $\pi/a$. 

As we have seen, a $\Lambda$~vertex vanishes in the limit $a\rightarrow
0$ unless momenta of order $\pi/a$ are important.  Thus, we expect that
a loop containing a $\Lambda$~vertex will receive a vanishing
contribution from the region of integration in which the magnitude of
the loop momentum is much smaller than $\pi/a$. Since the external
gauge-field momenta are assumed to be much smaller than $\pi/a$, one can
take the $a\rightarrow 0$ limit in this region simply by taking the
$a\rightarrow 0$ limits of the propagators and vertices, holding momenta
fixed. In this limit, propagators and naive single-gauge-field vertices
go over to continuum propagators and vertices, which are
$a$~independent, while multi-gauge-field naive vertices, Wilson
vertices, and $\Lambda$~vertices vanish as at least one power of $a$.
Furthermore, since the trace of an odd number of $\gamma$~matrices
vanishes, a $\Lambda$~vertex is always paired with a Wilson vertex or a
Wilson term in a propagator numerator. The volume of integration in this
region is independent of $a$. Thus, we conclude, that a loop that
contains a $\Lambda$~vertex receives a contribution from this region of
integration that vanishes as at least two powers of $a$ in the limit
$a\rightarrow 0$. 

Now we consider the region of integration in which the magnitude of the
loop momentum is of order $\pi/a$.  We can determine whether this is an
important region of integration by examining the sizes of the
propagators, vertices, and the domain of integration. (See, for example,
Ref.~\cite{bodwin-kovacs-schwinger} for further details.)  Away from its
pole at the origin, the propagator (\ref{prop-wilson}) is of order $a$.
Here, it is crucial that we have eliminated doublers; otherwise, there
would be poles in the propagator for components of the loop momentum
of order $\pi/a$.  An $n$-gauge-field-fermion vertex is of order $a^{n-1}$,
and a $\Lambda$-$n$-gauge-field-fermion vertex is of order $a^{n-1}$.
The domain of integration is of order $a^{-d}$ in $d$ dimensions. From
this it follows that the region in which the magnitude of the
fermion-loop momentum is of order $\pi/a$ gives a contribution of order
$a^{N_g-d}$, where $N_g$ is the number of external gauge fields. Note
that this result is independent of the number of
$\Lambda$~vertices.\footnote{Since we are concerned only with
infinitesimal gauge transformations, we need never consider the case of
more than one $\Lambda$~vertex.} We define the degree of divergence of a
loop to be 
\begin{equation}
D=d-N_g,
\label{deg-div}
\end{equation}
which corresponds to the expression in continuum field theory.  If the
loop is UV convergent, that is, if $D$ is negative, then the
contribution from the region in which the magnitude of the loop momentum
is of order $\pi/a$ vanishes as a power of $a$ in the limit
$a\rightarrow 0$. In this case, for a loop containing no
$\Lambda$~vertices, the contribution from the region in which the
magnitude of the loop momentum is much less than $\pi/a$ dominates.  One
can obtain the $a\rightarrow 0 $ limit of this contribution by replacing
the integrand with the continuum expression. The resulting integral is
UV convergent, and so one can extend the range of integration to
infinity with negligible error.  Hence, the $a\rightarrow 0$ limit of
this contribution is identical to the continuum amplitude. 

We conclude that a fermion loop containing a $\Lambda$~vertex gives a
vanishing contribution in the limit $a\rightarrow 0$, unless the degree
of divergence is non-negative.  Hence, for $d=4$, the gauge variations
that persist in the continuum limit arise only from loops involving a
$\Lambda$~vertex and four or fewer external gauge-field vertices. 

Using these same arguments, we can also conclude that a term in a loop
amplitude that is proportional to a Wilson mass or vertex gives a
contribution that vanishes as a power of $a$ in the limit $a\rightarrow
0$, unless the degree of divergence of the loop is non-negative. 
Furthermore, in the case a non-negative degree of divergence, the
dominant contribution comes from the region of integration in which the
loop momentum is of order $\pi/a$.  That is, the contribution takes the
form of a local interaction, with configuration-space size of the order
of the inverse of the lattice UV cutoff $\pi/a$. 

\subsection{Modifying the fermion determinant}
\label{sec:mod-det}

At this point we could attempt to restore the gauge symmetry by adding
renormalization counterterms to the theory.  Of course, no counterterm
can remove violations of the gauge symmetry that arise from the ABJ
anomaly. Partly because of the absence of full rotational symmetry on
the lattice, the number of possible counterterms is quite large.  In
addition to the usual rotationally invariant gauge-field wave-function
renormalization, there are counterterms corresponding to a gauge-field
mass, a rotationally non-invariant wave-function renormalization, and
rotationally invariant and non-invariant gauge-field--gauge-field
scattering amplitudes.  The tuning of all of these counterterms in a
lattice simulation would be awkward.  Fortunately, there is a trick that
can be used to implement the required counterterms automatically
\cite{lat90,dpf91,lat92}. Motivated by the fact that a theory with
vector-like couplings of the fermion to the gauge field is gauge
invariant, we will attempt to rearrange the fermion-loop amplitude so
that it looks like the loop amplitude for a vector-like theory. 

Consider an arbitrary fermion-loop amplitude. We can write the
projectors $P_L=(1-\gamma_5)/2$, which appear only in the naive
vertices, in terms of the unit matrix and $\gamma_5$ and expand the
expression for the amplitude. The result is a sum, each term of which
contains an even or an odd number of factors of $\gamma_5$. 

\subsubsection{The even-parity part}
\label{sec:even}

For those terms that contain an even number of $\gamma_5$'s, which we
call even-parity terms, we would like to move the factors of $\gamma_5$
together and use the identity $\gamma_5^2=1$ to eliminate them, thereby
obtaining the corresponding expression for a vector-like theory.  This
would amount to a simple algebraic manipulation, were it not for the
fact that $\gamma_5$ anticommutes with the naive terms in the
rationalized-propagator numerators and naive vertices, but commutes with
the Wilson terms in the rationalized-propagator numerators and Wilson
vertices.  We would obtain a result that is proportional to the
corresponding expression in a vector-like theory were we to treat
$\gamma_5$ as if it anticommuted with the Wilson terms in the
rationalized-propagator numerators and Wilson vertices. We will follow
this procedure.  Of course, the resulting expression will differ from
the original one, and we must account for this difference. However, the
difference is always proportional to a Wilson mass from a propagator
numerator or a Wilson vertex.  As we have demonstrated in
Section~\ref{sec:divergent}, a loop containing a Wilson mass or vertex
vanishes as at least one power of $a$ in the limit $a\rightarrow 0$,
unless the degree of divergence is non-negative, and then the
contribution corresponds to a local interaction.  Thus, such
contributions have the form of renormalization counterterms.  We can
drop them without affecting the nature of the theory: such a procedure
amounts merely to adding renormalization counterterms to the action and
choosing a particular tuning of the counterterm coefficients. Then, for
the terms in the original loop amplitude that contained an even number
of $\gamma_5$'s we obtain expressions that are proportional to the
corresponding expressions in a vector-like theory.  We now work out the
constants of proportionality. 

Consider first a contribution from a loop amplitude that contains at least
one naive vertex. We are interested only in manipulating the terms
containing an even number of $\gamma_5$'s. However, it is simplest to
work out the combinatorics by moving the complete projectors $P_L$ until
they stand next to each other, treating $\gamma_5$ as if it commuted
with all other factors in the amplitude. Each projector is separated by
$N$ propagators and $N$ vertices from another, and so, in the process of
moving one projector so that it is adjacent to another, the projector
flips from a $P_L$ to a $P_R$, but always winds up as a $P_L$ in the
end.  Since $P_L^2=P_L$, we have just one projector $P_L=(1-\gamma_5)/2$
when the process is finished.  The even-parity part of the amplitude
corresponds to the term $1/2$. Thus the even-parity part yields a
contribution that is exactly half the corresponding contribution in a
vector-like theory. 

Now consider a contribution from a loop amplitude that contains no naive
vertices.  In this case, there are no projectors $P_L$, the contribution
is entirely even in parity, and it is equal to the corresponding
contribution in a vector-like theory.  In order to combine it with the
even-parity parts of the contributions containing at least one naive
vertex, so as to obtain a complete vector-like amplitude, we must
discard half.  However, since the discarded piece contains no naive
vertices, it must contain at least one Wilson vertex. As we have already
argued, we can safely discard such a contribution, since that act
amounts to choosing a particular tuning of the coefficients of
renormalization counterterms. 

At the end of all of these manipulations, the even-parity part of a
fermion-loop amplitude yields a contribution that is half the
corresponding contribution in a vector-like theory.  The effective
action that one obtains by integrating over the fermion degrees of
freedom is, of course, given by the loop amplitudes, weighted by
$1/N_g$. Therefore, the effect of our manipulations is to replace the
even-parity part of the contribution to the effective action by one half
the effective action for a vector-like theory. Now, the lattice Dirac
operator ${\cal D}$, which is defined by 
\begin{equation}
a^d\sum_{x}\overline\psi(x){\cal D}\psi(x)=S_N+S_{NI}+S_W+S_{WI}
\label{dirac-op}
\end{equation}
has the property\footnote{This property also holds if one drops the 
gauging of the Wilson term $S_{WI}$ on the right side of 
(\ref{dirac-op}).} that 
\begin{equation}
{\cal D}|_{\gamma_5\rightarrow -\gamma_5}=\gamma_5{\cal D}^\dagger
\gamma_5.
\label{dirac-parity}
\end{equation}
Now, the effective action is given by 
\begin{equation}
S_{eff}=\ln(\det{\cal D}). 
\label{eff-action}
\end{equation}
Since $\det\gamma_5=1$, we see from (\ref{dirac-parity}) that 
\begin{equation}
\case 1/2 [S_{eff}\pm (S_{eff}|_{\gamma_5\rightarrow -\gamma_5})]
=\case 1/2 (S_{eff}\pm S_{eff}^\dagger).
\end{equation}
That is, the even-parity (odd-parity) part of the effective action is
the real (imaginary) part of the effective action.  Furthermore,
(\ref{eff-action}) implies that the real (imaginary) part of the
effective action corresponds to the magnitude (phase) of the fermion
determinant. 

{\it Therefore, we conclude that our manipulations amount to the
prescription that the magnitude of the chiral fermion determinant be
replaced by the square root of the fermion determinant for a vector-like
theory.}\footnote{There is no ambiguity in the sign of the square root.
We are identifying the square root with the {\it magnitude} of the
fermion determinant, and so we always take the positive sign.  The sign
ambiguity associated with the Witten anomaly \cite{witten} is carried by
the phase of the determinant.  Since the low-energy spectrum is
unchanged by our modifications of the determinant, the Witten anomaly is
unaffected. In particular, the Witten anomaly is absent in this lattice
implementation of the Standard Electroweak Model.} This prescription has
been discussed previously in the case of continuum theories
\cite{alvarez-gaume-pietra} and in the case of lattice theories
\cite{lat92,gockeler-schierholz}; an equivalent formulation involving
auxiliary fermion species has also been presented \cite{lat90,dpf91}. If
one adopts this prescription, then the magnitude of the fermion
determinant and, correspondingly, the real part of the effective action
have an exact gauge invariance. 

We note that, since these manipulations amount to the addition of
renormalization counterterms to the theory, they do not affect
unitarity.  This is obvious at the level of the action, since, in
Minkowski space, it is Hermitian even with the addition of counterterms.
It is also easy to see diagrammatically: a cut of a diagram can never
pass through a short-distance loop (momenta of order the UV cutoff),
because the on-shell conditions and energy-momentum conservation
constrain the components of the momenta of the cut lines to have
magnitudes much smaller than the UV cutoff. 

Of course, as we have already argued at the diagrammatic level, the
manipulations that we have made do not affect the low-energy behavior of
the theory.  It is easy to see this directly from the action.  The
even-parity part of the effective action generated by a fermion with
left-handed couplings to the gauge field is equal to one-half the effective
action generated by two fermions, one with left-handed couplings and one 
with right-handed couplings. The continuum limit of the action for such 
a complement of fermions is given by 
\begin{eqnarray}
&&\lim_{a\rightarrow 0}\sum_x [\overline\psi_1(x){\cal D}\psi_1(x)
+\overline\psi_2(x)({\cal D}|_{\gamma_5\rightarrow -\gamma_5})
\psi_2(x)]\nonumber\\
&&=\sum_x [\overline\psi(x)(\partial\cdot\gamma
+igA\cdot\gamma)\psi(x)
+\overline\psi_1(x)\partial\cdot\gamma P_R\psi_1(x)
+\overline\psi_2(x)\partial\cdot\gamma P_L\psi_2(x)],
\label{low-energy}
\end{eqnarray}
where $\psi=P_L\psi_1+P_R\psi_2$.
Here, in taking the continuum limit, we have assumed that the momenta
associated with the Fourier transforms of the fields are all fixed to be
much less than the UV cutoff, so that one can take the ``naive''
$a\rightarrow 0$ limit of operators.  We conclude that the even-parity
part of the effective action goes, at low momentum and in the  continuum
limit, to one half the effective action generated by a fermion with
vector-like couplings to the gauge field, plus non-interacting degrees
of freedom. 

\subsubsection{The odd-parity part}
\label{sec:odd}

Now we turn to the terms in the loop amplitude that contain an odd
number of $\gamma_5$'s, which we call the odd-parity part.  The
manipulations of the preceding section, which bring $\gamma_5$'s
together and use $\gamma_5^2=1$ to eliminate them, can never succeed in
converting the odd-parity parts to a vector-like amplitude: there will
always be one $\gamma_5$ left over in the end.  Thus, we must deal in
another way with the violations of the gauge symmetry in the odd-parity
parts that persist in the limit $a\rightarrow 0$. 

Let us specialize, for the moment, to four dimensions. As we have seen
in Section~\ref{sec:divergent}, the gauge variations that are
nonvanishing as $a\rightarrow 0$ are contained in the fermion-loop
amplitudes involving one $\Lambda$~field and four or fewer gauge
fields. Then, one can see that the nonvanishing gauge variations
correspond to the ABJ anomaly. An explicit calculation is presented in
Appendix~\ref{app:anomaly}. Here we give a general argument that the
gauge variations are zero, provided that one chooses a theory in which
the complement of physical fermions satisfies the anomaly-cancellation 
condition 
\begin{equation} 
{\rm Tr}\, (T_a\{T_b,T_c\})=0. 
\label{anom-cancel}
\end{equation} 

As we have argued in Section~\ref{sec:divergent}, a loop containing a
$\Lambda$ vertex receives a nonvanishing contribution in the limit
$a\rightarrow 0$ only from the region of integration in which the
magnitude of the loop momentum is of order $\pi/a$.  That means that the
nonvanishing gauge variations all have the form of local interactions.
In four dimensions, the odd-parity, local operators of dimension four or
less that are invariant under lattice rotations and involve a
$\Lambda$~field and 
gauge fields are of the form ${\rm Tr}\, [\Lambda
\epsilon_{\mu\nu\rho\sigma} A_\mu A_\nu A_\rho A_\sigma]$ and ${\rm Tr}\,
[\Lambda \epsilon_{\mu\nu\rho\sigma}(\partial_\mu A_\nu) A_\rho
A_\sigma]$, or ${\rm Tr}\,[\Lambda \epsilon_{\mu\nu\rho\sigma}(\partial_\mu
A_\nu) (\partial_\rho A_\sigma)]$.  These all vanish if the 
anomaly-cancellation condition (\ref{anom-cancel}) is satisfied.  There
remains the possibility that subleading contributions from this region
of integration could give rise to violations of gauge invariance that
vanish as powers of $a$. However, there are no lattice-rotationally
invariant, odd-parity, local operators of dimension five involving a
$\Lambda$~field and gauge fields.  Hence, the violations of gauge
invariance from the region of integration in which the magnitude of the
loop momentum is of order $\pi/a$ vanish at least as $a^2$ in the limit
$a\rightarrow 0$. 

Similar arguments show that, in two dimensions, the gauge variations of
the odd-parity part of a loop also vanish as $a^2$ in the limit
$a\rightarrow 0$, provided that the anomaly is cancelled. In two
dimensions one can achieve cancellation of the anomaly in a non-trivial
theory by introducing left-handed and right-handed fermions such that
the sum of ${\rm Tr}\,(T_a T_b)$ for the left-handed fermions is equal
to the sum of ${\rm Tr}\,(T_a T_b)$ for the right-handed fermions. 

We emphasize that, in contrast with the modified even-parity loop
amplitudes, the odd-parity loop amplitudes do not possess an exact gauge
invariance, even if (\ref{anom-cancel}) is satisfied.  There are
violations of the gauge symmetry that vanish only in the limit
$a\rightarrow 0$. We have just seen that such violations can arise from
the region of integration in which the fermion-loop momentum is of order
$\pi/a$.  In Section~\ref{sec:divergent}, we noted that violations of
gauge invariance can also arise from the region of integration in which
the magnitude of the fermion-loop momentum is much less than $\pi/a$,
even in UV-convergent diagrams.  In both of these cases, the violations
of gauge invariance vanish as $a^2$ in the limit $a\rightarrow 0$. 

The odd-parity amplitudes themselves are finite in the limit
$a\rightarrow 0$.  This follows from the fact that there are no
odd-parity renormalization counterterms involving only gauge fields. In
four dimensions, the lattice-rotationally invariant, odd-parity, local
operators of dimension four or less involving gauge fields have the
forms 
${\rm Tr}\, 
[ \epsilon_{\mu\nu\rho\sigma} A_\mu A_\nu A_\rho A_\sigma]$, ${\rm
Tr}\, [\epsilon_{\mu\nu\rho\sigma}(\partial_\mu A_\nu) A_\rho
A_\sigma]$, and ${\rm Tr}\,[\epsilon_{\mu\nu\rho\sigma} (\partial_\mu
A_\nu) (\partial_\rho A_\sigma)]$. When one symmetrizes under cyclic
permutations of the gauge fields, the first operator vanishes, and the
second and third operators are total derivatives.  It can be seen in a
similar fashion that corresponding operators in two-dimensional theories
vanish. Since the gauge variation of an odd-parity amplitude vanishes as
$a^2$ in the limit $a\rightarrow 0$, we can conclude that the deviation
of an odd-parity amplitude from a gauge-invariant expression also
vanishes as $a^2$. 

Finally, we mention that the analysis of the gauge variations of the
odd-parity parts of loops in this section does not depend on the gauging
of the Wilson term.  The analysis relies only on the power-counting
rules and the general structure of the local interactions, neither of
which are affected by the presence or absence of (\ref{wilson-gauge}) in
the action. 

\section{DYNAMICAL GAUGE FIELDS}
\label{sec:dynamical}

We wish to generalize the discussion of
Section~\ref{sec:background-gauge} to include the possibility that the
gauge fields are dynamical, rather than simply external background
fields.  The important distinction is that the gauge-field momentum can
now contain a loop momentum, and so its magnitude can range up to the
lattice cutoff $\pi/a$.  Now we can have divergent loop integrations
involving gauge-field propagators as well as fermion propagators, and
the results for the counting of powers of $a$ must be generalized from
those derived in Section~\ref{sec:background-gauge}. 

The even-parity parts of fermion loops can again be rendered exactly
gauge invariant by making use of the $\gamma_5$ trick of
Section~\ref{sec:even} to replace the fermion loop by one-half the
corresponding loop for a fermion with vector-like interactions with the
gauge field.  We have already seen that this replacement does not alter
the low-energy behavior of amplitudes. Therefore, it amounts to a change
of UV regulator, which is equivalent to the addition of counterterms to
the theory.  In the case of a background gauge field with momentum much
smaller in magnitude than the UV cutoff $\pi/a$, the required
counterterms were those generated by a single fermion loop.  In the
present case, counterterms can also be generated by multi-loop
subdiagrams, including loops involving gauge fields.  Fortunately, we do
not need to implement these counterterms explicitly: they are provided
automatically by modification of the fermion-loop amplitude. 

The case of the odd-parity parts of fermion loops is more complex and
requires some further analysis. 

\subsection{Counting powers of $a$}

We wish to study the gauge variations of the odd-parity parts of 
fermion loops in the limit $a\rightarrow 0$.  That is, we wish to study
the behavior of a diagram or a subdiagram containing exactly one
$\Lambda$~vertex in that limit.  As we argued in
Section~\ref{sec:background-gauge}, contributions involving a
$\Lambda$~vertex are suppressed by at least one power of $a$ in the
limit $a\rightarrow 0$ unless a momentum entering the $\Lambda$~vertex
has a magnitude of order $\pi/a$.  Thus, we wish to study the region of
integration in which the loop momenta have magnitudes of order $\pi/a$.
We might as well take all the loop momenta in a subdiagram to have
magnitudes of order $\pi/a$, since we can always study the case when
only a subset of the loop momenta have magnitudes of order $\pi/a$ by
considering a smaller subdiagram.  For purposes of the discussion in
this subsection only, we assume that the gauge field has been fixed to a
renormalizable gauge. 

Now we use the facts that, in the region in which all momenta have
magnitudes of order $\pi/a$, an $n$-gauge-field-fermion vertex is of
order $a^{n-1}$, a $\Lambda$-$n$-gauge-field-fermion vertex is of
order $a^{n-1}$, a fermion propagator is of order $a$, a gauge-field
propagator is of order $a^2$, an $n$-gauge-field vertex is of order
$a^{n-4}$, and each loop integration has a range of order $(1/a)^d$ in
$d$ dimensions.  From these facts, it is easy to see that a
single-particle-irreducible (1PI) diagram or subdiagram with $N_f$
external fermion legs, $N_g$ external gauge-field legs, and $L$ loops is
of order $a^{-D}$, where the degree of divergence $D$ is given
by\footnote{Ghost loops, which appear with certain choices of gauge, do
not affect these conclusions.} 
\begin{equation}
D=4-N_g-\case 3/2 N_f+L(d-4).
\label{deg-div-dyn}
\end{equation}

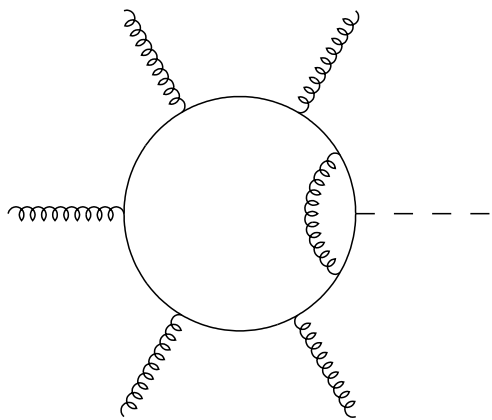
\begin{figure}
\caption{An example of a gauge variation whose odd-parity part is
nonvanishing in the continuum limit in four dimensions.  The circle
represents the fermion loop, the dashed line represents the
$\Lambda$~field, and the curly lines represent the gauge fields.} 
\label{fig:vertex-corr}
\end{figure}

Any 1PI subdiagram that contains a $\Lambda$~vertex and has a
non-negative degree of divergence can potentially lead to a violation of
the gauge symmetry that survives in the limit $a\rightarrow 0$. As we
have already argued, the even-parity parts of fermion loops in such a
subdiagram can be rendered exactly gauge-invariant by replacing the
fermion loop with one-half the corresponding loop for a fermion with
vector-like interactions with the gauge field.  However, in the case of
the odd-parity part of a loop, a $\Lambda$~vertex inside a radiative
correction can give a nonvanishing contribution in four dimensions. (An
example of such a contribution is shown in Fig.~\ref{fig:vertex-corr}.)
Hence, there are violations of the gauge symmetry in four dimensions. 

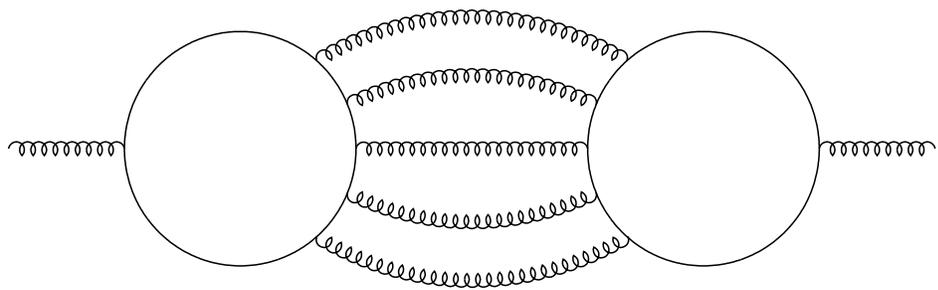
\begin{figure}
\caption{A contribution to the gauge-field self energy that leads to a 
violation of the gauge symmetry in four dimensions.  The violation 
arises from the odd-parity parts of the loops.}
\label{fig:gauge-self-energy}
\end{figure}

One might hope that it would be possible to restore the gauge symmetry
by tuning the limited number of renormalization counterterms that are
associated with divergent radiative corrections\cite{lat90,dpf91,lat92}.
Unfortunately, this turns out not to be the case.  For example, in four
dimensions, the diagram of Fig.~\ref{fig:gauge-self-energy} has an
overall degree of divergence $D=2$.  Thus, the contribution that arises
from the odd-parity parts of the fermion loops yields violations of the
gauge symmetry, even though the individual fermion loops have a negative
degree of divergence.  In particular, the diagram generates a
gauge-field mass, and so would require a mass counterterm, even if
gauge-field-mass generation has been eliminated at the one-loop level by
modifying the even-parity parts of loops as described in
Section~\ref{sec:even}. On examining other multi-loop diagrams, one
reaches the conclusion that that all possible renormalization
counterterms consistent with the cubic lattice symmetry appear. 

\subsection{The double-limit procedure}
\label{sec:double-limit}

We would like to restore the gauge invariance of the theory without
resorting to the tuning of counterterms.  If we could limit the momenta
in loops involving gauge fields to be much less than the fermion-loop UV
cutoff, then the arguments of Section~\ref{sec:background-gauge} would
apply. One way to achieve this is to introduce two different lattice
spacings, $a_g$ for the gauge field and $a_f$ for the fermion field and
take the limit $a_f\rightarrow 0$ with $a_g$ fixed before taking the
limit $a_g\rightarrow 0$.   Such a double-limit procedure is similar
in spirit to the UV regulator employed in proving the
anomaly-no-renormalization theorem \cite{adler-bardeen2}. A double-limit 
procedure has also been discussed previously in the context of
lattice theories
\cite{smit,gockeler-schierholz,anlreport,thooft,hsu,kronfeld}. The use
of a double limit along with the modification of the magnitude of the
fermion determinant has been discussed previously in
Refs.~\cite{gockeler-schierholz,anlreport}. 

\subsubsection{Interpolation of the gauge fields: general considerations}

In computing the double limit, we assume that the gauge-field links that
reside on the gauge-field lattice $U_{g\mu}$ are the dynamical
variables, {\it i.e.}, the variables over which one integrates in the
path-integral expressions for amplitudes.  These are the quantities that
appear in the pure gauge-field action (\ref{gauge-field-action}). The
interactions of the gauge fields with fermion fields are obtained by
inserting gauge-field links $U_\mu$, which reside on the fermion
lattice, into the fermion action as in (\ref{naive-action}),
(\ref{naive-gauge}), (\ref{wilson-action}), and (\ref{wilson-gauge}).
These gauge-field links that reside on the fermion lattice are not the
dynamical variables $U_{g\mu}$.  We must obtain them by an interpolation
of the dynamical gauge-field links. 

It is often convenient to discuss the interpolation in terms of the
gauge fields $A_\mu$, which are related to the plaquettes through
(\ref{plaquette}). One can use the Hamilton-Cayley theorem to express
the logarithm of an $m\times m$ group matrix (link), as a linear
combination of the unit matrix and the first $m-1$ powers of the matrix.
The ambiguity in the phase of the coefficients can be resolved by
requiring that matrices that are close to the unit matrix have
logarithms that are close to zero. This is equivalent to the requirement
proposed by 't~Hooft\cite{thooft} that the eigenvalues of $a_gA_{g\mu}$
and the eigenvalues of $a_fA_{f\mu}$ lie on the interval $(-\pi,\pi]$. 

For simplicity, we will assume that $a_g/a_f=R$ is an integer and that
the fermion lattice subdivides the gauge-field lattice, so that they
coincide every $R$ sites.  For each gauge-field-lattice site $y$, there
are $R^d$ fermion-lattice sites $x=y+m$, where $m$ is vector whose
components are integers satisfying 
\begin{equation}
0\leq m_\nu \leq (R-1). 
\end{equation}
We will also assume that, in the interpolation, the fermion-lattice
links $U_\mu$ depend only on the gauge-lattice links $U_{g\nu}$ that
form the edges of the surrounding hypercube.  That is, we assume that
the links $U_\mu(y+ma_f)$ depend only on the links
$U_{g\nu}(y+m_g(\nu)a_g)$, where $m_g(\nu)$ is a vector with integer
components satisfying 
\begin{equation}
\cases{
m_{g\rho}(\nu)=0&for $\rho=\nu$;\cr
m_{g\rho}(\nu)=0\hbox{\ or\ }1&for $\rho\neq\nu$.\cr
}
\label{m-prime}
\end{equation}
Similarly, the fermion-lattice fields $A_\mu(y+m_\mu a_f+\case 1/2 
a_{f\mu})$ depend only on the gauge-field-lattice fields 
$A_{g\nu}(y+m_g(\nu)a_g+\case 1/2 a_{g\mu})$.

The Fourier transform of the fermion-lattice field is given by
\begin{eqnarray}
\tilde A_\mu(l)&&=(a_f)^d\sum_xA_\mu(x+\case 1/2 a_{f\mu})
\exp[-i(x+\case 1/2 a_{f\mu})\cdot l]\nonumber\\
&&\equiv(a_g)^d\sum_y \exp[-i(y+\case 1/2 a_{g\mu})\cdot l]
\overline A_\mu(l,y+\case 1/2 a_{g\mu}),
\label{fourier-tx}
\end{eqnarray}
where 
\begin{equation}
\overline A_\mu(l,y+\case 1/2 a_{g\mu})=
\exp(\case i/2 a_{g\mu}\cdot l) R^{-d} \sum_m
A_\mu(y+ma_f+\case 1/2 a_{f\mu})
\exp[-i(ma_f+\case 1/2 a_{f\mu})\cdot l].
\label{a-bar}
\end{equation}
Note that, if $\overline A_\mu(l,y+\case 1/2 a_{g\mu})$ were equal to
$A_{g\mu}(y+\case 1/2 a_{g\mu})$, then $\tilde A_\mu(l)$ would be equal
to $\tilde A_{g\mu}(l)$, where
\begin{equation}
\tilde A_{g\mu}(l')=(a_g)^d\sum_y A_\mu(y+\case 1/2 a_{g\mu})
\exp[-i(y+\case 1/2 a_{g\mu})\cdot l']
\label{gauge-lattice-tx}
\end{equation}
is the Fourier transform of the field on the gauge-field lattice. 
We express the deviation of $\tilde A_\mu(l)$ from $\tilde A_{g\mu}(l)$ 
in terms of a ``regulating factor'' $F_\mu(l)$:
\begin{equation}
\tilde A_\mu(l)=F_\mu(l)\tilde A_{g\mu}(l).
\end{equation}

Many different interpolations of the gauge fields are possible. However,
if the interpolation is to lead to a gauge-invariant theory in the
double limit, then certain minimal requirements must be met: the
interpolation must lead to correct tree-level amplitudes in the continuum
limit, the interpolation must provide a UV cutoff of order $\pi/a_g$ on
gauge-field momenta, and the interpolation must relate a gauge
transformation of the fields on the gauge-field lattice to a gauge
transformation of the fields on the fermion lattice. We now enumerate a 
set of sufficient conditions for meeting these requirements.

{\def\theenumi{\alph{enumi}}
\begin{enumerate}

\item  {\it Locality.} The interpolation must be local in the sense that
gauge fields on the fermion lattice cannot depend on gauge fields on the
gauge-field lattice that are separated by an arbitrarily large number of
gauge-field-lattice sites.  If one were to employ a nonlocal
interpolation, then the gauge-field-fermion interactions would not go
to the continuum (local) form in the limit $a_g\rightarrow 0$.  The
interpolation need not be strictly local; it can depend on gauge fields
that are separated by a finite number of gauge-field-lattice sites.
However, a dependence of the interpolation on widely separated
gauge-field-lattice sites would lead to large order $a_g$ errors in the
limit $a_g\rightarrow 0$. We have assumed a local form for the
interpolation in (\ref{m-prime}). 

\item {\it Smoothness.}  We take as a smoothness requirement the
continuity of fields inside hypercubes on the gauge-field
lattice.\footnote{Such a criterion has been discussed in 
Refs.~\cite{thooft,kronfeld,gockeler-et-al}.} 
That is, we require that, for a given $y$, the fields $A_\mu(y+a_f 
m+\case 1/2 a_f)$ differ on adjacent fermion lattice sites by
quantities of order $a_f$. There can, depending on the interpolation, be
discontinuities along certain directions at the boundaries between the
gauge-field hypercubes.  However, the size of these discontinuities is
independent of $a_f$. 

The smoothness requirement leads to a UV cutoff on the gauge-field 
momentum, since it guarantees that the Fourier transform 
(\ref{fourier-tx}) vanishes as $a_f^n$ if $n$ components of $l$ are of 
order $\pi/a_f$.  We can see this by making use of the elementary 
properties of Fourier transforms.  Consider the one-dimensional Fourier 
transform
\begin{equation}
\tilde A_\mu(l_\nu)=a_f\sum_{x_\nu} A_\mu(x+\case 1/2 a_{f\mu})
\exp[-i(x_\nu +\case 1/2 a_{f\mu}) l_\nu]\qquad\hbox{(no sum over 
$\nu$)}.
\label{one-dim-tx}
\end{equation}
From (\ref{one-dim-tx}) it follows that
\begin{equation}
a_f\sum_{x_\nu}|\nabla_\nu^+A_\mu(x+\case 1/2 a_{f\mu})|^2
=\int_{-\pi/a_f}^{\pi/a_f}{dl_\nu\over 2\pi}
(4/a_f^2)\sin^2(\case 1/2 l_\nu a_{f\nu})
|\tilde A_\mu(l_\nu)|^2,
\label{deriv-tx}
\end{equation}
where
\begin{equation}
\nabla_\nu^\pm f(x)=\pm (1/a_f)[f(x\pm a_\mu)-f(x)]
\label{lat-deriv}
\end{equation}
are the forward and backward lattice derivatives. Smoothness requires
that the lattice derivative of the field $\nabla_\nu A_\mu$ be of order
$a_f^0$, except possibly at gauge-field-lattice hypercube boundaries,
where it may be of order $a_f^{-1}$. Since the number of boundaries does
not grow with decreasing $a_f$, the left side of (\ref{deriv-tx}) is at
most order $a_f^{-1}$.  This implies that, on the right side of
(\ref{deriv-tx}), $\tilde A_\mu(l_\nu)$ can be at most of order $a_f$
over a range of $l_\nu$ that is of order $\pi/a_f$. 

Smoother interpolations than we consider here lead to additional
suppression of the Fourier transform of the interpolated field at large
momentum. One can derive relations similar to (\ref{deriv-tx}), but
involving higher derivatives.  From these, it can be seen that, if,
along the $\nu$~direction, the $(r-1)$st derivative of $A_\mu(x+\case
1/2 a_{f\mu})$ is continuous and the $r$th derivative is continuous
except at hypercube boundaries, then $\tilde A_\mu(l_\nu)$ can be at
most order $a_f^{r+1}$ over a range of $l_\nu$ of order $\pi/a_f$.  It
should be noted, however, that as an interpolation becomes smoother, it
becomes increasingly less local, involving more widely separated sites
on the gauge-field lattice.  Therefore, such interpolations, in general,
increase the size of the order $a_g$ errors in the limit $a_g\rightarrow
0$. 

The smoothness requirement, coupled with locality, also guarantees that
one recovers the correct tree-level amplitudes in the continuum limit.
That is, it guarantees, that 
\begin{equation}
\lim_{a_gl\rightarrow 0}\tilde A_\mu(l)=\tilde A_{g\mu}(l).
\end{equation}
This follows immediately from the fact that, because of continuity,
\begin{equation}
\lim_{a_gl\rightarrow 0}\overline A_\mu(l,y+\case 1/2 a_{g\mu})=
R^{-d}\sum_m A_\mu(y+a_fm+\case 1/2 a_{f\mu})
\end{equation}
can differ from $A_{g\mu}(y+\case 1/2 a_{g\mu})$ only by a quantity of
order $Ra_f=a_g$. Here we are making use of the fact that the gauge
fields associated with the tree amplitudes are continuous on the
gauge-field lattice. 

Therefore, we conclude that the smoothness requirement leads to the 
properties
\begin{mathletters}
\label{f-props}
\begin{equation}
F_\mu(l)\sim a_f^n\qquad\hbox{if $n$ components of $l$ are of order 
$\pi/a_f$}
\label{f-cutoff-prop}
\end{equation}
and
\begin{equation}
F_\mu(l)\approx 1\qquad\hbox{for $l<<\pi/a_g$}.
\label{f-cont-prop}
\end{equation}
\end{mathletters}
As we have already mentioned, smoother interpolations result in additional
suppression of $F_\mu(l)$ when components of $l$ are large.  For
example, if the interpolation of $A_\mu$ is ``transversely continuous'',
{\it i.e.}, continuous along directions $\nu\neq\mu$ at the boundaries
of the gauge-field-lattice plaquettes, then there is an additional
power of $a_f$ on the right side of (\ref{f-cutoff-prop}) for each
component $l_\nu$ that is of order $\pi/a_f$. 

\item{\it Gauge Covariance.} We require that, for every gauge
transformation $\Lambda'$ of the gauge-field-lattice links $U_{g\mu}$,
the interpolation of the gauge-transformed links $U_{g\mu}^{\Lambda'}$
must yield a set of fermion-lattice links $U_\mu^\Lambda$, where
$\Lambda$ denotes a gauge transformation of the fermion-lattice links
$U_\mu$ \cite{gockeler-et-al}.  This requirement allows one to infer,
from the gauge invariance of the fermion sector of the theory on the
fermion lattice, that the complete theory on the gauge-field lattice is
gauge invariant. 

One might imagine that one could meet this gauge-invariance requirement
by fixing to a particular gauge before carrying out the interpolation.
However, gauge fixing is a nonlocal procedure and, therefore, violates the 
requirement that the interpolation be local.

\end{enumerate}}

The interpolations that we will consider have the property that the
gauge field $A_\mu$ is constant along fermion-lattice links $U_\mu$ that
lie along gauge-field lattice links $U_{g\mu}$. For these links, $A_\mu$
is chosen to be equal to $A_{g\mu}$.  This implies that 
\begin{equation}
U_{g\mu}(y)=\prod_{m_\mu}U_\mu(y+m_\mu a_f)
=[U_\mu(y)]^R.
\label{link-prod}
\end{equation}
In solving (\ref{link-prod}) for $U_\mu$, we choose the branch cut of
the $R$th root in accordance with the definition of the gauge fields
discussed earlier in this section.  That is, we take the branch cut such
that, if $U_{g\mu}$ is near unity, then $U_\mu$ is near unity. 

The property (\ref{link-prod}) is compatible with the gauge-covariance
requirement. In order to see that this is so, consider a gauge
transformation 
$\Lambda'(y)$ on the gauge-field lattice.  Each link $U_{g\mu}$ is
transformed according to 
\begin{equation}
U_{g\mu}(y)\rightarrow \exp[i\Lambda'(y)]
U_{g\mu}(y)\exp[-i\Lambda'(y+a_{g\mu})].
\label{gauge-tx-full}
\end{equation}
Thus, according to (\ref{link-prod}), the fermion-field links change 
as follows:
\begin{equation}
U_\mu(y+m_\mu a_f)=[U_{g\mu}(y)]^{1/R}\rightarrow 
\{\exp[i\Lambda'(y)]U_{g\mu}(y)\exp[-i\Lambda'(y+a_{g\mu})]\}^{1/R}.
\label{interp-tx}
\end{equation}
A gauge transformation $\Lambda$ on the fermion-lattice links that 
reproduces the right side of (\ref{interp-tx}) can be obtained by the 
following procedure.  First, set 
\begin{equation}
\Lambda(y)=\Lambda'(y)\qquad \hbox{for all $y$.}
\label{gauge-coinc}
\end{equation}
Then, each link $U_\mu(y+m_\mu a_f)$ can be brought into agreement with
the right side of (\ref{interp-tx}) by suitable choice of
$\Lambda(y+m_\mu a_f+a_{f\mu})$, where the choices can be made by a
sequential algorithm, starting at the first link and working toward the
last link.  At the last link, the choice of $\Lambda(y+m_\mu
a_f+a_{f\mu})=\Lambda(y+a_{g\mu})$ must not conflict with
(\ref{gauge-coinc}). However,
\begin{eqnarray}
&&\prod_{m_\mu}\{\exp[i\Lambda(y+m_\mu a_f)]U_\mu(y+m_\mu a_f)
\exp[-i\Lambda(y+m_\mu a_f+a_{g\mu})]\}\nonumber\\
&&=\exp[i\Lambda(y)]\prod_{m_\mu}[U_{g\mu}(y+m_\mu a_f)]
\exp[-i\Lambda(y+a_{g\mu})],
\end{eqnarray}
and so the choice of $\Lambda(y+a_{g\mu})$ that is required by 
(\ref{link-prod}) is 
\begin{equation}
\Lambda(y+a_{g\mu})=\Lambda'(y+a_{g\mu}),
\end{equation}
which agrees with (\ref{gauge-coinc}).

Recently, Shamir \cite{shamir} has pointed out that there is a potential
difficulty in maintaining the smoothness and the gauge covariance of the
interpolation procedure.  He has shown that the interpolating field
differs from a smooth field by a gauge transformation that is, in
general, topologically nontrivial and, hence, singular. These
difficulties do not appear in an Abelian theory with a non-compact
gauge-field action.  It is possible that they might be avoided by fixing
to a suitable gauge on the gauge-field lattice. However, this issue has
yet to be resolved.  In the analyses to follow, we indicate those parts
of the arguments that may be affected by these considerations. 

\subsubsection{An Abelian interpolation}

As an example, let us consider an interpolation that satisfies the 
required properties in the case of an Abelian theory.  In an Abelian 
theory, the gauge transformation (\ref{gauge-tx-full}) is equivalent to 
\begin{equation}
A_{g\mu}(y+a_{g\mu}/2)
\rightarrow A_{g\mu}(y+a_{g\mu}/2)+(1/ag)[\Lambda'(y)
-\Lambda'(y+a_{g\mu})].
\label{abelian-gauge-tx}
\end{equation}
If the interpolation of the $\Lambda'$-dependent part of
(\ref{abelian-gauge-tx}) has a vanishing lattice curl, then it can be
written as the lattice gradient of a potential on the fermion lattice.
Then (\ref{abelian-gauge-tx})) is equivalent to a gauge transformation
on the fermion-lattice fields (of the same form as
(\ref{abelian-gauge-tx})). It is easy to see that a simple linear 
interpolation of the gauge field \cite{gockeler-et-al} has this property.  
Hence, it is gauge covariant under infinitesimal gauge transformations
(although not under the large gauge transformations of
Ref.~\cite{shamir}). To be explicit, one takes 
\begin{eqnarray}
A_\mu(y+\case 1/2 a_{f\mu}+ma_f)&&\,=\sum_{m_g(\mu)}
A_{g\mu}(y+\case 1/2 a_{g\mu}+a_g m_g(\mu))\nonumber\\
&&\qquad\times\prod_{\nu\neq\mu}
[(1-m_\nu/R)(1-m_{g\nu}(\mu))+(m_\nu/R)m_{g\nu}(\mu)].
\label{abelian-interp}
\end{eqnarray}
Clearly, this interpolation satisfies the locality and smoothness
requirements.
We have, for this interpolation,
\begin{equation}
\overline A_\mu(l,y+\case 1/2 a_{g\mu})=A_{g\mu}(y+\case 1/2 a_{g\mu})
{\sin(\case 1/2 a_f l_\mu R)\over R\sin(\case 1/2 a_f l_\mu )}
\prod_{\nu\neq\mu} \left[{\sin^2(\case 1/2 a_f l_\nu R)\over 
R^2\sin^2(\case 1/2 a_f l_\nu )}\right],
\end{equation}
which implies that the regulating factor is given by 
\begin{equation}
F_\mu(l)={\sin(\case 1/2 a_f l_\mu R)\over R\sin(\case 1/2 a_f l_\mu )}
\prod_{\nu\neq\mu} \left[{\sin^2(\case 1/2 a_f l_\nu R)\over 
R^2\sin^2(\case 1/2 a_f l_\nu )}\right].
\end{equation}
We see explicitly that the properties (\ref{f-props}) hold, 
as expected from our general arguments.

\subsubsection{A non-Abelian interpolation}
\label{sec:non-abelian}

In the case of non-Abelian gauge fields, simple linear interpolations of
the sort discussed in the last section do not satisfy the
gauge-covariance requirement. However, 't~Hooft \cite{thooft} has
proposed a more intricate interpolation method that does. Here we
discuss a variant of 't~Hooft's method that was suggested by Hern\'andez
and Sundrum \cite{hernandez-sundrum}. 

The first step in the method is to fix the interpolation for 
fermion-lattice links that lie along gauge-field-lattice links according 
to (\ref{link-prod}).  As we have already shown, this step is consistent 
with the gauge-covariance requirement.

The next step is to determine the interpolation for the fields $A_\mu$
that lie on the two-dimensional surface of an elementary plaquette,
where here $\mu$ is either one of the two  directions that define the
plaquette. The interpolation is given by the field configuration that
minimizes the two-dimensional action for a pure gauge-field theory on
the fermion lattice,\footnote{This action is given by
(\ref{gauge-field-action}), but in two dimensions and on the fermion
lattice.} subject to the boundary conditions on the fields on the links
bounding the plaquette.  To obtain a unique solution to the minimization
condition, one must fix the gauge. A convenient choice is the
two-dimensional Lorentz gauge 
\begin{equation}
\sum_{\mu=1}^d\nabla_\mu^- A_\mu=0.
\label{lorentz-gauge}
\end{equation}

One can argue that the solution is unique as follows.  The minimization
condition implies that the field configurations satisfy the gauge-field
equations of motion. If we neglect terms of higher order in $a_f$, then the 
equation of motion is
\begin{equation}
(\nabla_\mu^- -igA_\mu) F_{\mu\nu}=0,
\end{equation}
where
\begin{equation}
F_{\mu\nu}=\nabla_\mu^+ A_\nu-\nabla_\nu^+ A_\mu-ig[A_\mu,A_\nu],
\end{equation}
and we have rescaled the fields by $g$. In the Lorentz gauge, the
equation of motion becomes 
\begin{equation}
\nabla_\mu^- \nabla_\mu^+  A_\nu
-ig\nabla_\mu^- [A_\mu,A_\nu]+g^2(i[A_\mu,A_\nu])^2=0. 
\label{eq-motion}
\end{equation}
If one sets $g=0$ in (\ref{eq-motion}), then one recovers
Laplace's equation, which, with the given boundary conditions, has a
unique solution.  One can obtain a solution to all orders in $g$ by
iteration, treating the order $g$ and order $g^2$ terms as source terms
and using the solution to Laplace's equation as a starting point. Hence,
in the continuum limit, the interpolated field configuration that is
continuously connected to the $g=0$ solutions is unique. 

In order to see that the gauge fields derived through this interpolation
procedure satisfy the smoothness requirement, suppose the opposite: that
a gauge field has a discontinuity. Then, for at least one point $x$, the
first term on the left side of (\ref{eq-motion}) is of order $a_f^{-2}$,
whereas the remaining terms are of order $a_f^{-1}$ or smaller. (Here we
are assuming that the interpolated gauge field is bounded, which may not
be true in the presence of singularities of the type discussed by Shamir
\cite{shamir}.) Therefore, in the case of a discontinuous gauge field,
the equations of motion cannot be satisfied in the continuum limit, and
one concludes that the gauge field does not satisfy minimization
criterion in the continuum limit. 

In four dimensions, there are two more steps in the interpolation
method. The third step is to determine the fields inside the cubes
bounded by the elementary plaquettes.  One does this by seeking a field
configuration that minimizes the three-dimensional pure gauge-field
action, subject to the boundary conditions along the elementary
plaquettes and the three-dimensional Lorentz-gauge condition.  The last
step is to determine the fields inside the four-dimensional hypercubes
bounded by the three-dimensional cubes.  One minimizes the
four-dimensional pure gauge-field action, using the fields on the cubes
as boundary conditions and fixing to the four-dimensional Lorentz gauge.
It is easy to see, by generalizing the preceding arguments, that these
last two steps result in fields that satisfy the smoothness requirement.

Finally, there is the question of whether this interpolation method
satisfies the gauge-covariance requirement.  Suppose that we have
obtained a field configuration on the fermion lattice by the
interpolation method.  Then suppose that we make a gauge transformation
on the gauge-field lattice.  The links bounding the elementary
plaquettes will be changed in value, and a re-application of the
interpolation procedure will result in a new field configuration on the
fermion lattice.  We wish to show that this new field configuration can
be obtained by a gauge transformation on the fermion lattice of the
original fermion-lattice field configuration.  Here, we paraphrase the
argument presented in Ref.~\cite{hernandez-sundrum}. 

We have already shown that there is a gauge transformation that does
this for the gauge fields that lie on the links bounding the
elementary plaquettes on the gauge-field lattice. Such a gauge
transformation will not, in general, leave the gauge fields that lie
inside the plaquettes in the two-dimensional Lorentz gauge. However, we
can always make a gauge transformation on the {\it interior} of a
plaquette that returns the fields to the Lorentz gauge, without changing
the fields on the links that bound the plaquette. Similarly, we can find
a gauge transformation on the interior of a three-dimensional cube that
returns the fields inside the cube to the three-dimensional Lorentz
gauge and a gauge transformation on the interior of a four-dimensional
hypercube that returns the fields inside the hypercube to the
four-dimensional Lorentz gauge.  Since the pure gauge-field actions are
invariant under these transformations, the resulting configuration still
satisfies the minimization criteria. Hence, it is identical to the field
obtained by applying the interpolation method to the gauge-transformed
gauge-field-lattice links.  Here we are assuming the uniqueness
of the interpolated field configuration.

\subsubsection{The Feynman rules}

By considering the Fourier transform of the lattice action, one can
easily derive the Feynman rules for the double-limit procedure. 

The Feynman rules for the gauge-field propagators and vertices are the
same as those for a theory with lattice spacing $a_g$.  Momenta in
propagators and vertices range from $-\pi/a_g$ to $\pi/a_g$, and momentum
is conserved modulo $2\pi/a_g$.  Hence, pure gauge-field loop
integrations range from $-\pi/a_g$ to $\pi/a_g$. 

The Feynman rules for fermion propagators, gauge-field-fermion vertices,
and $\Lambda$-fermion vertices are determined by considering the Fourier
transform of the fermionic part of the action. Momenta in propagators
and vertices range from $-\pi/a_f$ to $\pi/a_f$ and momentum is
conserved modulo $2\pi/a_f$.  Hence, pure fermionic loop integrations
range from $-\pi/a_f$ to $\pi/a_f$. 

When a gauge-field line attaches to a fermion line, one must 
consider the effect of the interpolation in working out the Fourier 
transform of the gauge field on the fermion lattice, as in 
(\ref{fourier-tx}).  The interpolation introduces
a regulating factor $F_\mu(l)$ for each connection of a gauge-field 
line to a fermion line.  The gauge-field momentum $l$, which
appears in the Fourier transform of the gauge field on the fermion 
lattice (\ref{fourier-tx}), can be written as 
\begin{equation}
l=l'+(2\pi/a_g) q,
\label{mom-decomp}
\end{equation}
where $q$ takes on values from $(-R+1)/2$ to $(R-1)/2$ in integer steps and
$-\pi/a_g\leq l' <\pi/a_g$. We can think of the integration over $l$
from $-\pi/a_f$ to $\pi/a_f$ as an integration over $l'$ from $-\pi/a_g$
to $\pi/a_g$ and a sum over $q$ from $(-R+1)/2$ to $(R-1)/2$. There is
an integration over $l'$ and a sum over $q$ for each attachment of a
gauge-field or $\Lambda$~line to a fermion line. The quantity $l'$ may be 
interpreted as the gauge-field momentum variable in the Fourier
transform of the gauge field on the gauge-field lattice
(\ref{gauge-lattice-tx}). Only $l'$ appears in gauge-field propagators
and pure gauge-field vertices; they are insensitive to the value of $q$
because, as can be seen from (\ref{gauge-lattice-tx}) they are periodic,
with period $2\pi/a_g$. In a Feynman diagram, integrations over
variables of the type $l'$ are constrained by the fact that the total of
the gauge-field momentum, including the variables of the type $l'$, is
conserved, modulo $2\pi/a_g$, in {\it every} propagator and vertex.
Thus, the gauge-field-momentum variables, including those of the type
$l'$, can be re-organized, in the usual way, into independent
loop momenta, which range from $-\pi/a_g$ to $\pi/a_g$, and external
momenta. In general, the fermion propagators, gauge-field-fermion
vertices, and $\Lambda$~vertices depend on the value of $q$, as well as
on the value of $l'$.  The sums over variables of the type $q$ are
constrained only by momentum conservation, modulo $2\pi/a_f$, along each
fermion line. Aside from this constraint, there is an independent sum
over $q$ for each attachment of a gauge-field line to a fermion line. 

Using these Feynman rules and (\ref{f-cont-prop}), we see that, in the
limit $a_g\rightarrow 0$, for momenta much less than the cutoff
$\pi/a_g$, the Feynman rules for the fermion become the continuum
Feynman rules.  Therefore, we recover the required low-energy behavior
of the tree-level amplitudes. 

\subsubsection{Counting powers of $a_f$}
\label{sec:counting-a-f}

In this section we will demonstrate, for an open fermion line or for the
odd-parity part of a closed fermion line, that contributions that arise
when gauge-field (or $\Lambda$-field) momenta of order $\pi/a_f$ enter
the line vanish in the limit $a_f\rightarrow 0$ with $a_g$ fixed. We
call momenta of order $\pi/a_f$ ``large'' momenta. In the arguments 
to follow, we assume that the even-parity parts of fermion loops have
been modified as in Section~\ref{sec:even} to render them exactly gauge
invariant.  One consequence of this assumption is that all of the gauge
variations must arise from the odd-parity parts of loops.  The argument
that we present holds in two and four dimensions. We proceed by counting
the powers of $a_f$ associated with a contribution in which large
gauge-field (or $\Lambda$-field) momenta enter a fermion line. 

In the initial discussion, we assume that the fermion-loop momentum
associated with a closed fermion line is not large. Since momentum is
conserved, modulo $2\pi/a_f$, along a fermion line, if one gauge-field
momentum entering a fermion line is large, at least one other
gauge-field momentum entering a fermion line must be large. We
assume, initially, that exactly two gauge-field momenta entering a 
fermion line are large.

Powers of $a_f$ arise from the fermion propagators, gauge-field-fermion
vertices, and $\Lambda$~vertices through which the large momentum flows.
It is easy to see, by making use of the power-counting rules of
Section~\ref{sec:divergent}, that the minimum number of factors of $a_f$
arises if the large gauge-field momentum flows through at most one
fermion propagator. 

Inverse powers of $a_f$ can arise from the sum over variables of the
type $q$ in (\ref{mom-decomp}).  If two gauge-field momenta entering a
fermion line are large, there is only one independent sum, the other sum
being constrained by momentum conservation. The range of the sum
contributes a factor of order $R\sim a_f^{-1}$ for each component of
the momentum that is large. 

There is a regulator factor $F$ associated with each of the points at
which the two large momenta enter the fermion line. From
(\ref{f-cutoff-prop}), we see that each regulator factor contributes a
factor $a_f$ for each component of the momentum that is large. Hence,
the minimum number of powers of $a_f$ is obtained by taking only one
component of the momentum to be large. 

By way of illustration, let us consider the case in which the large
gauge-field momentum flows through exactly one fermion propagator.  As
we have already noted, this case gives the minimum number of powers of
$a_f$. The fermion propagator contributes a factor of order $a_f^1$. The
large momentum also flows through two gauge-field-fermion vertices or a
gauge-field-fermion vertex and a $\Lambda$~vertex.  The
gauge-field-fermion vertices contribute factors of order $a_f^0$ and the
$\Lambda$~vertex contributes a factor of order $a_f^{-1}$. Hence, the
propagators and vertices contribute a factor of order $a_f^1$ in the
amplitude and $a_f^0$ in the gauge variation. If we take one component
of the gauge-field momentum to be large, the range of the sum over $q$
gives $a_f^{-1}$ and the regulator factors give $a_f^2$.  We conclude
that, in this example, the contribution to the amplitude from the
factors associated with the large gauge-field momenta is of order
$a_f^2$.  The contribution to the gauge variation is larger---of order
$a_f^1$. This is a consequence of the fact that the large momentum
associated with the $\Lambda$~field contributes an additional
dimensionful factor of $1/a_f$ to the gauge variation. Since the
contributions to the amplitude itself from this momentum region vanish
as $a_f^2$, we can still conclude that the amplitude differs from a
gauge-invariant expression by terms of order $a_f^2$. 

Now let us relax the assumption that only two of the gauge-field momenta
entering the fermion line are large.  For each additional large
momentum, there is at least one factor $a_f$ for the propagators and
vertices through which it flows, a factor $a_f^{-1}$ for the 
associated sum over $q$, and a factor $a_f$ from the associated 
regulator factor.  Hence, contributions involving more than two large 
gauge-field momenta are suppressed by at least one additional power of 
$a_f$.

We can also relax the assumption that the fermion-loop momentum 
associated with a closed fermion line is not large.  Suppose that the 
loop momentum is large.  Then, the entire contribution of the loop, 
including the sums over variables of the type $q$ and the regulator 
factors, arises from short distances and can be expressed in terms of local 
operators on the gauge-field lattice.

Consider first the case of loops containing gauge variations
($\Lambda$~vertices).  All of the gauge variations arise from odd-parity
loops.  As we have already discussed in Section~\ref{sec:odd}, the
lattice-rotationally invariant, odd-parity, local operators of dimension
$d$ or less involving a $\Lambda$~field are of the form of the ABJ
anomaly. (In the present case, continuum derivatives must be replaced by
lattice derivatives on the gauge-field lattice, since we are really
discussing the effective theory on the gauge-field lattice.) These all
vanish if the anomaly-cancellation condition (\ref{anom-cancel}) is
satisfied. There are no such operators of dimension $d+1$. Hence, the
contributions to the gauge variations from the regions of integration in
which both the gauge-field momenta and the fermion-loop moment are large
are of order $a_f^2$, possibly times logarithms of $a_f$. 

Now consider the odd-parity parts of loop amplitudes.  Recalling our
arguments of Section~\ref{sec:odd} (and again replacing continuum
derivatives by derivatives on the gauge-field lattice), we note that the
lattice-rotationally invariant, local, odd-parity operators of dimension
$d$ or less involving only gauge fields all vanish under Bose
symmetrization. Furthermore, there are no lattice-rotationally invariant,
odd-parity, local operators of dimension $d+1$ involving only gauge
fields. Hence, the contributions to the odd-parity loop amplitudes from
the regions of integration in which both gauge-field momenta and the
fermion-loop moment are large are of order $a_f^2$, possibly times
logarithms of $a_f$. 

Finally, we consider the even-parity parts of loop amplitudes. Because
the even-parity parts of loops are exactly gauge invariant, only
gauge-invariant local operators can contribute.  There {\it is} a
lattice-rotationally invariant, gauge-invariant, Bose-symmetric operator
of dimension $d$, namely, the one that renormalizes the gauge-field
wave function. Hence, there could, in principle, be contributions, in
which large gauge-field momenta flow into the even-parity parts of
loops, that go as $a_f^0$, possibly times logarithms of $a_f$.  Of
course, we need not show that such contributions vanish in order to
establish the gauge invariance of the double-limit procedure.
Furthermore, their behavior is no worse than that of the even-parity
parts of fermion loops in the absence of large gauge-field momenta,
which is also logarithmic in $a_f$. 

We must also consider the possibility that, in a Feynman diagram,
inverse powers of $a_f$ could arise from a fermion loop other than the
fermion line under consideration, and thereby lead to contributions from
regions of large gauge-field momenta that are nonvanishing as
$a_f\rightarrow 0$. We have already seen that such inverse powers of
$a_f$ cannot arise when gauge-field momenta entering the loop are large
and or when both gauge-field momenta and the fermion-loop momentum are
large.  The local-operator argument given for the latter case also
applies when only the fermion-loop momentum is large. Therefore, no
inverse powers of $a_f$ can arise from a fermion loop.

Let us summarize these results.  We have found that, in the double
limit, contributions in which a large gauge-field momentum enters a
fermion loop containing a gauge variation vanish as $a_f$ times
logarithms of $a_f$. This result, combined with the analysis of
Section~\ref{sec:background-gauge}, allows us to conclude that the
odd-parity parts of fermion loops can be rendered gauge invariant by
taking the double limit and by requiring the fermion to be in a
representation of the gauge group that satisfies the
anomaly-cancellation condition. We assume that the even-parity parts of
fermion loops have been rendered exactly gauge invariant by replacing
them with one-half the corresponding loop for a fermion with a
vector-like coupling to the gauge field. Therefore, we have achieved our
goal of making all the amplitudes in the theory gauge invariant.  We
have also found that contributions associated with the odd-parity parts
of loops are finite in the limit $a_f\rightarrow 0$.  This implies that
the phase of the fermion determinant is finite in this limit.
Furthermore, we have seen that the contributions in which a large
gauge-field momentum enter the odd-parity part of a fermion loop vanish
as $a_f^2$, possibly times logarithms of $a_f$. This result, together
with the analysis of Section~\ref{sec:odd}, implies that the phase of the
fermion determinant differs from a gauge-invariant expression by terms
of order $a_f^2$, possibly times logarithms of $a_f$, in the limit
$a_f\rightarrow 0$. 

It should be noted that the detailed power-counting rules we have
presented in this subsection are specific to interpolations of the gauge
fields that are discontinuous in at least one direction at the
boundaries of the gauge-field hypercubes. One might devise smoother
interpolations in which the gauge fields (or their higher derivatives)
are continuous. For such interpolations, the regulating factor
$F_\mu(l)$ and, hence, the contributions to the amplitudes and gauge
variations would be suppressed by additional factors of $a_f$ when
gauge-field-fermion-loop momenta are of order $\pi/a_f$. 

It may be useful to contrast our results with those of
Ref.~\cite{hernandez-sundrum}.  In that work, the authors make the
additional assumption that the interpolation is transversely continuous.
(That assumption is valid for the interpolations that we have
presented.)  They are then able to show that the all the contributions
in which a large gauge-field momentum enters a fermion loop are
suppressed by powers of $a_f$.  Their proof applies to the even-parity
parts of loops, as well as to the odd-parity parts of loops and to loops
containing $\Lambda$~vertices.  They conclude, as we do, that
contributions in which large gauge-field momenta enter the odd-parity
parts of loops vanish as $a_f^2$.  However, they also conclude that
gauge variations vanish as $a_f^2$.  This last result seems to be at
odds with our explicit example. 

\subsubsection{Options for computing the determinant}
\label{sec:imp-opt}

In the last section we demonstrated that there exists a satisfactory
procedure for computing the fermion determinant.  There are actually
several variants of this procedure that one can employ, and some may be
more efficient than others in practical calculations.  We now discuss
some of these computational options. 

Once one has replaced the magnitude of the fermion determinant with the
square root of the determinant for a fermion with vector-like couplings
to the gauge field, the magnitude of the fermion determinant has an
exact gauge invariance. Therefore, one can evaluate the modified
magnitude of the determinant without employing the double-limit
procedure, and still obtain a gauge-invariant result. That result will
be equivalent to the one obtained through the doubling-limit procedure,
since the effective action is unique, aside from gauge-invariant
counterterms, which can always be absorbed into a redefinition of the
coupling constant. 

There are several advantages in calculating the magnitude of the fermion
determinant without making use of the double-limit procedure. There
is the obvious advantage that one would not be faced in a numerical
simulation with the computational burden of taking the limit
$a_f\rightarrow 0$ for each gauge-field configuration. Another advantage
follows from the fact that, in four dimensions, the magnitude of the
determinant is divergent in the limit $a_f\rightarrow 0$. The divergence
arises from the diagram with two external gauge fields, which generates
the logarithm of $a_f$ that is associated with the gauge-field
wave-function renormalization.  In the double-limit procedure, one
would need to add a wave-function-renormalization counterterm, which has
the effect of replacing $\ln a_f$ with $\ln a_g$, to obtain the correct
renormalization of the gauge-field-fermion coupling and to obtain a
finite result. This counterterm can be determined from a one-loop
calculation, since radiative corrections to the fermion loop with two
external gauge fields are suppressed in the limit $a_f\rightarrow 0$.
However, it is simpler to bypass the double limit altogether in the case
of the magnitude of the determinant. 

One must, of course, make use of the double-limit procedure in
computing the phase of the determinant. Fortunately, in two and four
dimensions, the phase is finite in the limit $a_f\rightarrow 0$,
because, as we have seen, there are no odd-parity, Bose-symmetric
renormalization counterterms. 

There is one advantage in using the double-limit procedure to
compute the magnitude of the determinant. The vector-like gauge symmetry
of the magnitude of the determinant does not preclude the generation of
a mass for the fermion field.  In general, the unrenormalized fermion
mass will be nonzero. However, it is easy to see that fermion
self-energy diagrams are suppressed in the double-limit procedure. 

In the absence of the double-limit procedure, one must tune a
counterterm ({\it i.e.,} the hopping parameter $\kappa$) to make the
renormalized mass of the fermion with vector-like couplings vanish. In
practical terms, this procedure is somewhat tricky because we wish to
maintain the positivity of Wilson-Dirac determinant, so that its square
root is real. Of course, it is well known, from studies of theories with
vector-like interactions, how to determine the critical value of the
hopping parameter, $\kappa_{\rm critical}$, at which the renormalized
fermion mass vanishes. There are several procedures at one's disposal.
For example, one can use the vanishing of mass corrections to the
Ward-Takahashi identities, the vanishing of the Goldstone-boson (meson)
mass, or the first occurrence of a zero eigenvalue of the Wilson-Dirac
operator as definitions of $\kappa_{\rm critical}$. These approaches are
equivalent in the infinite volume limit. In determining $\kappa_{\rm
critical}$ by any of these methods, one averages over an ensemble of
gauge configurations.  A given gauge configuration may yield a value of
$\kappa_{\rm critical}$ that differs from the ensemble average.
Therefore, if one fixes $\kappa$ to be slightly below the
ensemble-average value of $\kappa_{\rm critical}$, one may encounter
``exceptional'' gauge-field configurations, such that the lowest
eigenvalue of the Dirac operator is negative and the fermion determinant
is negative. On the other hand, we expect an average of the determinant
over an ensemble of gauge configurations to be positive in the
thermodynamic limit of an infinite number of configurations.  To the
extent that the thermodynamic limit is equivalent to the infinite-volume
limit, the number of ``exceptional'' gauge-field configurations should
become vanishingly small as the volume is taken to infinity. 

We note that it is straightforward to reduce the size of the
gauge-variant contributions that arise from the odd-parity parts of
fermion loops in the region of integration in which the fermion-loop
momentum and gauge-field momenta all have magnitudes much less than
$\pi/a_f$.  These contributions are a consequence of order $a_f$
deviations of the tree-level lattice fermion action from the tree-level
continuum fermion action.  Such deviations are easily removed by
employing an improved tree-level action \cite{wetzel,eguchi-kawamoto}.
To reduce the size of the gauge variations that arise from the
low-momentum region, it is necessary only to improve the Wilson term in
the tree-level action.  

Similarly, one can eliminate the leading gauge-variant contributions
that arise from the odd-parity parts of loops in the region of
integration in which the gauge-field momenta entering a loop are large,
but the fermion-loop momentum itself is small.  As we have seen, these
contributions arise from subdiagrams in which the factors along the
fermion line are the same as in a one-loop fermion self-energy diagram. 
In particular, the leading contribution comes from the terms
corresponding to a fermion-mass renormalization. Mass generation is
precluded if the action is invariant under a constant shift of the
fermion field \cite{golterman-petcher}.   If we drop the gauging of the
Wilson term (\ref{wilson-gauge}), then the action exhibits this
symmetry.\footnote{It is easy to understand diagrammatically why mass
generation cannot occur. If the Wilson term is not gauged, then there
are no Wilson vertices, only naive vertices. Each of these contains a
$\gamma$~matrix and a factor $P_L$. Consider a fermion-self-energy
diagram. A Wilson mass from a rationalized propagator numerator vanishes
when sandwiched between two naive vertices, because of the projectors
$P_L$.  The remaining terms in the propagator numerators yield
contributions with an odd number of $\gamma$~matrices, and so they don't
have the form of a mass term.} In the case of the odd-parity parts of
loops, all of the arguments in both this section on dynamical gauge
fields and in Section~\ref{sec:background-gauge} on background gauge
fields are independent of whether the gauging of the Wilson term
(\ref{wilson-gauge}) is retained or not. Hence, we are free to drop the
gauging of the Wilson term in computing the phase of the determinant.
(In computing the magnitude of the determinant, one must retain the
gauging of the Wilson term in order to maintain the vector-like gauge
symmetry.) 

Unfortunately, the two improvement schemes that we have mentioned are
of no use unless one can also reduce the size of the violations of gauge
invariance that arise from the regions of integration in which both
gauge-field momenta and fermion-loop momenta are of order $\pi/a_f$.
This probably would require the use of smoother interpolations, which,
as we have already argued, ultimately require nonlocality 
and lead to increased errors of order $a_g$. 

Although the violations of gauge invariance vanish as powers of $a_f$, a
sufficiently large gauge transformation could make the coefficient of
the gauge variation impractically large for numerical work.  Therefore,
it is probably advantageous to fix the interpolating field to a smooth
gauge, such as one of the renormalizable gauges.  Then one would at
least avoid the spurious, large, ``pure gauge'' contributions to the gauge
field that are known to arise from UV divergences.

\section{MATRIX ELEMENTS OF FERMION OPERATORS}
\label{sec:operators}
Since a chiral-fermion action (for example, the sum of
(\ref{naive-action}), (\ref{naive-gauge}), (\ref{wilson-action}), and
(\ref{wilson-gauge})) is not invariant under gauge transformations, if
one computes matrix elements of operators involving fermion fields
straightforwardly using such an action, the result is not, in general,
gauge invariant. In this section, we discuss a method for computing matrix
elements of fermion operators that yields a gauge-invariant result. The
method that we present is related, but not identical, to the approach
that we used in computing the fermion determinant. 

In analyzing the matrix elements of fermion operators, we assume that
any fermions in the initial and final states have been removed by the
LSZ reduction.  We also assume that the total number of $\psi$'s is
equal to the total number of $\overline \psi$'s, so that the fermion
operators can be Wick contracted to form interacting propagators. 

\subsection{General procedure}

We begin by employing the $\gamma_5$~trick of Section~\ref{sec:even} to
move all the factors $P_L$ to the end points of the interacting fermion
propagators, treating $\gamma_5$ as if it anticommuted with all Wilson
masses and vertices.  If each interacting propagator's end points are
separated by a fixed amount in configuration space, then there is no
fermion-loop UV divergence associated with the propagator. In this case,
the rearrangement changes the expression by terms of order $a_f$ and by
terms corresponding to the renormalization counterterms associated with
radiative corrections to the propagators and operator vertices.  If the
interacting propagator's endpoints are separated by a distance that
vanishes as $a\rightarrow 0$, then there is a fermion-loop UV divergence
associated with the propagator.  In this case, the rearrangement also
changes the expression by terms corresponding to the renormalization
counterterms associated with the fermion loop. Once we have completed
this rearrangement, all of the factors $P_L$ are associated with the
fermion operators. Of course, $P_L^2=P_L$, and so there is at most one
such factor associated with the left side and one such factor associated
with the right side of each operator. 

If the operators themselves are independent of the gauge field, then the
modified matrix element is exactly gauge invariant, since the fermion
now has only vector-like interactions with the gauge field along its
propagators. Therefore, in this case, we can compute the modified matrix
element without recourse to the double-limit procedure. 

If an operator involves gauge fields, for example, through a
gauge-covariant derivative, then, with the modification that we have
described, the even-parity part of the expression associated with that
operator is still exactly gauge invariant, but the odd-parity part is
not.  Therefore, we can compute the even-parity part without making use
of the double-limit procedure.  For the-odd parity part, we must
invoke the double-limit procedure to ensure gauge invariance. 
If an interacting propagator's endpoints are separated by a distance that
vanishes as $a\rightarrow 0$, then nonvanishing gauge variations can
arise from the associated fermion-loop divergence. In this case, as was
discussed in Section~\ref{sec:odd}, we must also impose the
anomaly-cancellation condition (\ref{anom-cancel}) in order to ensure
gauge invariance.\footnote{Since we have applied the $\gamma_5$ trick
here to the odd-parity part as well as to the even-parity part, the
anomaly takes on a somewhat different form than in
Appendix~\ref{app:anomaly}. However, the conclusion---that the gauge
variations in the presence of a background field can be removed by
imposing the anomaly cancellation condition (\ref{anom-cancel})---is
unchanged.} 

The power-counting arguments that we have given previously also apply to
the operator matrix elements.  In particular, we expect the violations
of gauge invariance arising from odd-parity operator loops to vanish as
$a_f^1$, and we expect the deviations of the odd-parity loops from a
gauge-invariant expression to vanish as $a_f^2$. 

\subsection{Example: violation of baryon-number conservation}
\label{sec:baryon-number}

As an example of the procedure for computing matrix elements of operators
involving fermion fields, let us consider the matrix element of the
baryon-number current 
\begin{equation}
J_\mu^B(x)=\overline\psi^B(x)\gamma_\mu\psi^B(x) 
\end{equation} 
in the presence of dynamical gauge fields plus an external source of
background gauge-field quanta. We assume that $\psi^B$ is part of a
larger column vector $\psi$ such that the gauge group of the complete
field $\psi$ satisfies the anomaly-cancellation condition
(\ref{anom-cancel}), but the subgroup associated with $\psi^B$ does not.

A matrix element of $J_\mu^B$ is given by a weighted average over 
gauge-field configurations of 
\begin{equation}
F_\mu=\sum_x{\rm Tr}\, \gamma_\mu S_{\rm chiral}^B(x,x),
\label{J-gauge-var}
\end{equation}
where $S_{chiral}^B(x,x')$ is the interacting baryon propagator, with
configuration-space end points $x$ and $x'$. The subscript ``chiral''
indicates that the interactions of the baryons with the gauge field are
left handed. Now, $F_\mu$ is gauge variant.  However, we can modify the
definition of the matrix element so as to render it gauge invariant.  We
apply the $\gamma_5$~trick of Section~\ref{sec:even} to move all of the
projectors $P_L$ in $S_{\rm chiral}^B$ on the right side of
(\ref{J-gauge-var}) to the factor $\gamma_\mu$.  The terms that we
discard in this procedure all vanish in the limit $a_f\rightarrow 0$ or
have the the forms of renormalization counterterms. The result is that
$F_\mu$ is replaced by 
\begin{equation}
\tilde F_\mu=\sum_x{\rm Tr}\, \gamma_\mu P_L S_{\rm vector}^B(x,x),
\label{J-gauge-inv}
\end{equation}
where $S_{\rm vector}^B$ is the interacting propagator for baryons with
vector-like couplings to the gauge field. The expression
(\ref{J-gauge-inv}) has an exact (vector-like) gauge invariance. 
Consequently, we can compute it without recourse to the double-limit
procedure. 

Now $\tilde F_\mu$ corresponds to the matrix element of a left-handed
baryon current 
\begin{equation}
\tilde J_\mu^B(x)=\overline\psi^B(x)\gamma_\mu P_L\psi^B(x)
\end{equation}
in a theory in which the baryons have vector-like interactions with the
gauge field.  As is well known, in four dimensions, in a theory with
vector-like couplings, $\tilde J_\mu^B$ is not conserved: its divergence
is given by the ABJ anomaly, which is nonzero in the presence of
background gauge fields with nonzero winding number. Thus, we have
recovered the familiar result that, once one has added such
renormalization counterterms as are required to render its matrix
elements gauge invariant, the baryon-number current is not conserved 
\cite{dugan-manohar}. 

Of course, one could also compute the violation of baryon-number
conservation directly, by examining amplitudes that have unequal numbers
of incoming and outgoing baryons.  Such amplitudes can be computed in
the standard way by considering the contributions to the path integral
of the zero modes of the Dirac operator \cite{thooft-instantons}.  As we
have argued in Section~\ref{sec:even} (see, in particular, 
(\ref{low-energy})), the manipulations of the fermion determinant that
we advocate do not affect the low-energy modes in the continuum
limit.  Therefore, the lattice and continuum calculations yield the same 
result.

\section{BEYOND PERTURBATION THEORY}
\label{sec:nonpert}

The analyses that we have presented so far have been given in terms of
weak-coupling perturbation theory.  In this section, we will argue that,
in the presence of an arbitrary background gauge field, the perturbation
expansions for the fermion determinant and interacting fermion
propagators actually determine these quantities completely, except at
the zero modes of the Dirac operator. This is not to imply that one can
analyze the complete theory through the use of perturbative techniques.
The gauge-field sector of the theory, of course, exhibits effects that
are not amenable to a perturbative analysis. 

Throughout this section, we will assume that the gauge-field
configuration defined on the gauge-field lattice (and implicitly on the
fermion lattice) is bounded.  Of course, there is no universal bound
that applies to all of the gauge-field configurations in the path
integral. Therefore our conclusions may not hold when one sums over all
configurations. Another potential loophole arises from the fact that,
configuration by configuration, the gauge fields on the fermion-field
lattice may become unbounded because of singularities in the 
interpolating field of the type discussed by Shamir \cite{shamir}. 

\subsection{Finite volume and fixed lattice spacing}

In the arguments to follow, the convergence properties of the
perturbation series are crucial. Ultimately, we wish to study these
properties in the case of infinite volume and in the limit
$a_f\rightarrow 0$. However, it is illuminating to consider first the
behavior of the perturbation series for the somewhat simpler case of
finite volume and fixed lattice spacing. 

We begin by noting that the determinant of the lattice Dirac operator 
${\cal D}$ can be written as 
\begin{eqnarray}
\det{\cal D}&&=\det[\partial +({\cal D}-\partial)]
=\det\partial\,\det[1+(1/\partial)({\cal D}-\partial)]\nonumber\\
&&=\det\partial\,\exp\{{\rm Tr}\,\ln[1+(1/\partial)({\cal 
D}-\partial)]\},
\label{det}
\end{eqnarray}
where $\partial$ is the free Dirac operator (${\cal D}$ evaluated at
$g=0$).  The perturbation expansion for the effective action
$\ln(\det{\cal D})$ is obtained by expanding the logarithm in
(\ref{det}) in powers of $g$. 

As an intermediate step in analyzing the perturbation series, let us
examine the series in $1/\partial({\cal D}-\partial)$, introducing a
parameter $\zeta$ as the coefficient of $1/\partial({\cal D}-\partial)$
in (\ref{det}). At fixed lattice spacing in a finite volume, ${\cal D}$
and $\partial$ are just finite matrices.  Therefore, the logarithm can
be considered to be a matrix-valued function with matrix argument.
Furthermore, its expansion in powers of $\zeta (1/\partial)({\cal
D}-\partial)$ has a finite radius of convergence. Let $\lambda$ be an
eigenvalue of $1/\partial({\cal D}-\partial)$.  Then the radius of
convergence of the logarithm as a matrix-valued function of $\zeta$ is
$1/|\lambda_{max}|$, where $\lambda_{max}$ is the $\lambda$ with the
largest magnitude.  There is a branch-point singularity in the
matrix-valued function whenever $\zeta\lambda=-1$. 

Now, $({\cal D}-\partial)$ is an analytic function of $g$ through the
link variables $U$. Since $({\cal D}-\partial)$ vanishes as at least one
power of $g$ as $g\rightarrow 0$, the perturbation series has a finite
radius of convergence in $g$. The branch points at $\zeta\lambda=-1$
correspond to isolated branch points in the complex $g$~plane.
Consequently, one can determine $\det{\cal D}$ almost everywhere in the
complex $g$ plane by analytic continuation in $g$.  Of course, there are
ambiguities because of the cuts that arise from the branch points.
However, the ambiguity associated with a cut has no effect on the
determinant, since it leads to shifts of the argument of the exponential
by $2\pi in$, where $n$ is an integer.\footnote{In computing the square
root of the determinant of the Wilson-Dirac operator, we choose
$\kappa<\kappa_{\rm critical}$.  This implies that we are to the right
of the cut in $\det{\cal D}=\exp({\rm Tr}\,\ln{\cal D})$, and so there
is no ambiguity in the square root.} The branch points themselves
correspond to zero modes of the Dirac operator.  As we have argued in
Section~\ref{sec:even}, the procedure that we use to rearrange the
determinant leaves the zero modes unaffected; they are given, in the
limit $a_f\rightarrow 0$, by the zero modes of the continuum Dirac
operator. 

Similarly, we can write the interacting propagator as 
\begin{equation}
{\cal D}^{-1}=[\partial +({\cal D}-\partial)]^{-1}
=\partial^{-1}[1+({\cal D}-\partial)(1/\partial)]^{-1}.
\label{prop}
\end{equation}
The expansion of the right side of (\ref{prop}) in powers of $({\cal
D}-\partial)$ has a finite radius of convergence.  Therefore, the
perturbation expansion of ${\cal D}^{-1}$ in powers of $g$ has a finite
radius of convergence.  By analytic continuation, the perturbation
series determines the interacting propagator everywhere except at the
zero modes of the Dirac operator. 

\subsection{Infinite volume and the limit $a_f\rightarrow 0$}

Now let us take up the infinite-volume case.  Here it is most convenient 
to examine the convergence properties of the perturbation series, using 
the momentum-space Feynman rules.  We are ultimately interested in the 
limit $a_f\rightarrow 0$.

In order to demonstrate that our perturbative analyses hold for
arbitrary $g$, we need to prove two properties: that the perturbation
series for the effective action (logarithm of the fermion determinant)
and the interacting fermion propagator have finite radii of convergence,
and that one can take the limit $a_f\rightarrow0$ term by term in the
perturbation series. To prove the first property, we need to show only
that the perturbation series is absolutely convergent. To prove the
second property, we must show that the perturbation series is uniformly
convergent as $a_f\rightarrow 0$.  We will demonstrate this by showing
that the series can be majorized.  That is, we will show that for every
$a_f$ in a neighborhood of $a_f=0$, the absolute value of each term in
the perturbation series is bounded by an $a_f$-independent series that
converges.  Thus, the proof of the uniform convergence of the series
also demonstrates the absolute convergence of the series. We will assume
that the first few terms in the perturbation series of order $g^d$ or
less have been removed, so that we do not have to deal with individual
terms in the determinant that are divergent as $a_f\rightarrow 0$. 
Obviously, subtracting a finite number of terms does not affect the
convergence of the series. 

First we analyze the region of integration in which all the gauge-field
momenta, and the fermion-loop momentum in the case of the effective
action, have magnitudes much less than $\pi/a_f$. Consider the
contribution to a term of order $g^n$ that contains only
single-gauge-field-fermion vertices ${\cal V}^{(1)}$.  The magnitude of
each vertex is bounded by an $a_f$-independent constant times $g$. We
can obtain a bound on the magnitude each fermion propagator by dropping
the Wilson term and replacing $(1/a_f)\sin(p_\mu a_f)$ by a finite
constant of order unity times $p_\mu$.  Thus, the magnitude of each
propagator is bounded by an $a_f$ independent constant times $1/|p|$.
Since we are assuming, in the case of contributions to the effective
action, that the fermion momentum is much less than $\pi/a_f$, the
volume of the integration is an $a_f$-independent constant. Thus, each
such contribution to the interacting fermion propagator is bounded by
$C(gA/k)^n$, and each such contribution to the effective action is
bounded by $(1/n)C(Ag/k)^n$, where $C$ is an $a_f$-independent constant,
$A$ is the maximum magnitude of the gauge field\footnote{Here we are
assuming that the gauge-field configuration in momentum space is
bounded.  In fact, the gauge field may be singular in momentum space. 
However, if the gauge field is bounded in configuration space, then
these singularities are integrable. Hence, one could eliminate any such
singularities by smearing the momentum-space gauge field over a small
fraction of the range of the gauge-field momentum integration.},
and $k$ is the minimum of the magnitudes of the gauge-field momenta.
Here, we assume that the momentum of the gauge field is cut off in the
infrared by physical effects or by application of an explicit infrared
regulator. We also assume that one can neglect the regions of
integration in which sums of gauge-field momenta nearly vanish or, in
the case of the interacting propagator, sums of gauge-field momenta and
the fermion momentum nearly vanish.\footnote{Suppose that we constrain
$r$ momentum integrations so that each component of momentum has a range
of size $\epsilon$ relative its unconstrained range. There are
$n!/[(n-r)!r!]$ ways to do this.  The volume of integration of each of
the $r$ momenta is reduced by a factor $\epsilon^d$. At most $r$
propagators are enhanced by a factor $1/\epsilon$. Therefore, the net
effect of constraining momenta is to multiply the bounds we have
obtained by $(\epsilon^{d-1}+1)^n\leq C^n$, where $C$ is an
$a_f$-independent constant.} 

Suppose that we include the possibility of multi-gauge-field-fermion
vertices.  The effect of these is to replace propagator factors by
powers of $a_f$.  Therefore, we can bound any propagator factor by
$[C_1a_f+(C_2/k)]$, where $C_1$ and $C_2$ are $a_f$-independent
constants. This implies that the contributions to the interacting
propagator are bounded by $(Ag)^n[C_1a_f+(C_2/k)]^n$ and the
contributions to the effective action are bounded by
$(1/n)(Ag)^n[C_1a_f+(C_2/k)]^{n}$. Thus, we see that, for $g$ small
enough, these contributions are bounded by the terms in a convergent
geometric series that is independent of $a_f$. 

Now consider the region of integration in which some of the gauge-field
momenta are of order $\pi/a_f$.  As we have seen in
Section~\ref{sec:counting-a-f}, such contributions are suppressed by
powers of $a_f$. If a gauge-field momentum of order $\pi/a_f$ passes
through a fermion propagator, then the propagator is bounded by a
constant times $a_f$. Thus, we can again bound the propagator factors by
$[C_1a_f+(C_2/k)]$. There are additional powers of $a_f$ from the
regulating factors $F$ associated with the vertices. Otherwise, the
bounds on vertices are unchanged.  The powers of $a_f$ in the regulating
factors more than compensate for inverse powers of $a_f$ associated with
the ranges of the sums over the gauge-field-momentum variables $q$ in
(\ref{mom-decomp}). Therefore, the contributions to the interacting
propagator and the effective action are again bounded by
$(Ag)^n[C_1a_f+(C_2/k)]^n$ and $(1/n)(Ag)^n[C_1a_f+(C_2/k)]^n$,
respectively. For $g$ small enough, these quantities are, in turn,
bounded by the terms in a convergent geometric series that is
independent of $a_f$. 

Finally, we consider contributions to the effective action from the
region of integration in which the fermion-loop momentum is of order
$\pi/a_f$. We see from (\ref{deg-div}) and the surrounding discussion
that, for gauge-field momenta with magnitudes much less
than $\pi/a_f$, such contributions are bounded by an $a_f$-independent
constant times $(Ag)^na_f^{n-d}$.  The argument of the preceding
paragraph shows that contributions from gauge-field
momenta of order $\pi/a_f$ do not change this bound. Again, for $g$
small enough, the contributions are bounded by the terms in an
$a_f$-independent, convergent geometric series. 

We conclude that the perturbation series for the interacting propagator
and the effective action have finite radii of convergence and are
uniformly convergent in the limit $a_f\rightarrow 0$.  Therefore, the
perturbation series determine the propagator and the fermion determinant
by analytic continuation, except at singularities. Furthermore, we can
take the limit $a_f\rightarrow 0$ term by term. In this limit, the
singularities correspond to the zero modes of the continuum Dirac
operator. 

Therefore, the conclusions that we have reached through a perturbative
analysis of the fermion determinant and interacting propagator apply for
arbitrary $g$. In particular, we can conclude that, in the continuum
limit, the prescriptions we have given for computing the fermion
determinant and the matrix elements of fermion operators give the
correct low-energy amplitudes and yield gauge-invariant expressions. 

\section{SUMMARY AND DISCUSSION}
\label{sec:summary}

We have presented a general procedure for constructing gauge-invariant
lattice formulations of theories of chiral fermions interacting with
gauge fields.  The procedure involves three key ingredients: (1)~the
fermions must be in an anomaly-free representation of the gauge group;
(2)~one must replace the magnitude of the fermion determinant with the
square root of the determinant for a fermion that has vector-like
couplings to the gauge field, but that is otherwise identical to the
original fermion; (3)~one must implement the gauge-field action on a
lattice with spacing $a_g$ and the interacting fermion-field action on a
lattice with spacing $a_f$, define a suitable interpolation of the gauge
field to the fermion-field lattice, and take the limit $a_f\rightarrow
0$ before taking the limit $a_g\rightarrow 0$.\footnote{A typical
fermion action is given by the sum of (\ref{naive-action}),
(\ref{naive-gauge}), (\ref{wilson-action}), and (\ref{wilson-gauge}). 
The corresponding action for a fermion with vector-like couplings is
obtained by setting $P_R=P_L=1$.} In four dimensions, all three of these
conditions are required to ensure the gauge invariance of the
formulation.  In this procedure, the magnitude of the determinant is
exactly gauge invariant.  The gauge variations of the phase of the
determinant vanish as $a_f^1$, and the deviations of the phase of the
determinant from a gauge-invariant expression vanish as $a_f^2$,
possibly times logarithms of $a_f$. (We note that the result of
Ref.~\cite{hernandez-sundrum} for the power behavior of the gauge
variations seems to differ from the one derived in this paper.) 

We have also presented a closely related method for defining, in a
gauge-invariant fashion, matrix elements of fermion operators in chiral
theories. As was shown in Section~\ref{sec:baryon-number}, the
application of this method to the baryon-number current leads to the
familiar conclusion that that current is not conserved. 

The analysis of these methods is couched in weak-coupling perturbation
theory. In analyzing the properties of a UV regulator, of which the
lattice is an example, we are concerned with the behavior of the theory
near the cutoff.  Hence, one might hope, in the case of asymptotically
free theories, that the perturbation expansion would be a reliable guide
to that behavior. 

Furthermore, as we have argued in Section~\ref{sec:nonpert}, in the
presence of a given gauge-field configuration, the perturbation series
defines the interacting fermion propagator and the fermion determinant
everywhere except at zero modes of the Dirac operator.  The convergence
of the series is uniform in $a_f$, so that one can analyze the continuum
limit term by term. Hence, the methods for computing the determinant and
propagator are valid in the presence of a nonperturbative gauge-field
configuration.  We have not addressed the issue of the summation over
gauge-field configurations outside of the perturbative analysis. 

Shamir \cite{shamir} has presented an argument that potentially
undermines these analyses. He observes that, if an interpolation of the
gauge field is gauge covariant, then the interpolating field is related
to a smooth field by a gauge transformation that is, in general,
topologically nontrivial.  Hence, the interpolating field may possess
singularities. Such singular fields violate the smoothness requirement
for gauge fields on the fermion-field lattice that was used in the
power-counting analyses of Section~\ref{sec:counting-a-f} and also
violate the assumption of the boundedness of the gauge fields that was
made in Section~\ref{sec:nonpert}.  It is possible that one might avoid
these difficulties by fixing to a suitable gauge on the gauge-field
lattice. However, this is an open question. 

Putting aside questions of principle, it is not yet clear that the
procedure presented will be tractable in practical numerical
calculations.  The obvious stumbling block is the double-limit
procedure for $a_f$ and $a_g$, which could lead to computing
requirements that are much greater than in the case of a single
lattice-spacing limit. 

In computing the {\it magnitude} of the fermion determinant, one has two
distinct options.  One can apply the double-limit procedure. Then one
must tune a counterterm that renormalizes the gauge-field wave function
in order to keep the magnitude of the determinant finite in the limit
$a_f\rightarrow 0$ and to obtain the correct renormalization of the
gauge-field-fermion coupling. The coefficient of this counterterm is
readily computed in perturbation theory, since it is generated only by
the diagram with a single fermion loop and two external gauge fields. 

On the other hand, the magnitude of the fermion determinant is exactly
gauge invariant, once one has replaced it with the square root of the
determinant for a fermion with vector-like interactions.  Therefore, one
can compute the magnitude of the determinant by taking $a_f=a_g$.  Since
a vector-like gauge symmetry does not preclude the generation of a
fermion mass, one must also tune a mass counterterm (hopping parameter),
so as to keep the fermion massless.\footnote{The diagrams that generate
fermion masses are suppressed in the double-limit procedure, and so no
mass counterterm is required in that case.}  (In practice, it may be a
challenging problem to approach the critical value of the hopping
parameter in such a way that the positivity of the determinant is
maintained. See the discussion in Section~\ref{sec:imp-opt}.) In this
single-limit procedure, all other renormalization counterterms can be
absorbed into a redefinition of the coupling constant.  Hence, only the
fermion mass and the coupling constant need be tuned in taking the
continuum limit. 

It seems possible that one would need to compute only the magnitude of
the fermion determinant in updating gauge-field links, computing the
phase of the determinant as an expectation value once equilibrated
lattices had been generated.  If this turns out to be the case, then the
use of a single-limit procedure for the magnitude of the determinant
would result in an even greater relative reduction of the computing
time. 

In computing the phase of the fermion determinant one {\it must} employ
the double-limit procedure. This computation is mitigated somewhat in
two and four dimensions by the fact that, owing to the absence of
odd-parity counterterms in an anomaly-free theory, the phase is actually
finite in the limit $a_f\rightarrow 0$.  Therefore, one can carry out a
straightforward extrapolation to obtain the limit. 

One source of error in the extrapolation is easily reduced.  As we have
seen in Section~\ref{sec:dynamical}, order-$a_f^2$ deviations of the
phase of the determinant from a gauge-invariant expression arise from
the region of integration in which the gauge-field momenta and the
fermion-loop momentum associated with a given fermion loop are much
smaller in magnitude than $\pi/a_f$.  In this region, the deviations
from the limiting result come from the deviations of the tree-level
lattice action from the tree-level continuum action. The order in $a_f$
of these deviations can readily be increased through the use of improved
actions \cite{wetzel,eguchi-kawamoto}.  Similarly, one can eliminate the
order-$a_f^2$ gauge-variant contributions to the phase of the
determinant that arise from the region of integration in which
gauge-field momenta are large and the associated fermion-loop momentum
is small. One can accomplish this by dropping the gauging of the Wilson
term (\ref{wilson-gauge}) in computing the phase of the fermion
determinant (but not the magnitude). Then there is a symmetry under
constant shifts of the fermion field \cite{golterman-petcher} that
precludes the generation of fermion-mass terms, which give the largest
gauge-variant contributions. 

Unfortunately, such improvement programs are of limited utility, since
errors also arise from the region of integration in both gauge-field
momenta and fermion-loop momenta are of order $\pi/a_f$.  As we showed
in Section~\ref{sec:counting-a-f}, when one uses an interpolation in
which the gauge field is discontinuous along least one direction at the
boundaries of the gauge-field lattice hypercubes, these errors are of
order $a_f^2$. The use of a smoother interpolation, in which the gauge
fields (or higher derivatives) are continuous, would, in general,
suppress these errors by additional factors of $a_f$. However, such
interpolations are necessarily less local. In general, the order $a_g$
errors increase as one increases the distance on the gauge-field lattice
between the sites that enter in the interpolation. 

Although gauge-variant contributions ultimately vanish as
$a_f\rightarrow 0$, the presence of large, ``pure gauge'' contributions
in gauge-field configurations might make the approach to that limit
problematic in numerical work.  It is probably sensible, therefore, to
fix the interpolating field to a smooth gauge, such as one of the
renormalizable gauges, to ensure at least that the known, spurious,
``pure gauge'' contributions are absent. 

In testing the ideas of this paper in numerical simulations, it would be
most efficient, computationally, to consider two-dimensional theories.
Then, anomaly cancellation can be achieved by introducing both left- and
right-handed fermions, such that the sum of ${\rm Tr}\,(T_a T_b)$ for the
left-handed fermions is equal to the sum of ${\rm Tr}\,(T_a T_b)$ for the
right-handed fermions \cite{montvay}.  Strictly speaking,
two-dimensional theories do not require the double-limit procedure.
That is because, as can be seen from (\ref{deg-div-dyn}), the only
divergent subdiagram is a fermion loop with exactly two external gauge
fields; there are no divergent subdiagrams containing gauge-field 
propagators. However, the odd-parity part
of a fermion loop with two external gauge fields is zero by virtue of
the anomaly-cancellation condition (\ref{anom-cancel}). Therefore, the
violations of gauge invariance that arise from the odd-parity parts of
fermion loops vanish in limit $a_f=a_g\rightarrow 0$. Nevertheless, one
could use the two dimensional theories as a testing ground for methods
of extrapolating to the limit $a_f\rightarrow 0$ with $a_g$ fixed.  One
could check the gauge invariance of the fermion determinant and also
compare the results for various physical quantities, such as the mass
spectrum, with analytic results. 

More stringent tests of the methods presented here could be obtained in
four dimensions. Again, one could test the convergence of the
extrapolation to $a_f=0$ and the gauge invariance of the determinant.
Also, in weak coupling, one could compare results for physical
quantities in the standard electroweak model with calculations in
weak-coupling perturbation theory. 

It is clear that the fermion determinant we have described corresponds
to a complex effective action.  This is a general property of chiral
gauge theories that would be expected to hold regardless of the lattice
formulation chosen:  the effective action receives imaginary
contributions that are {\it independent} of the UV regularization from
finite odd-parity parts of fermion loops. It remains an open question as
to whether one can devise practical means for handling such complex
actions in numerical simulations. 

\acknowledgements

I would like to thank Maarten Golterman, Eve Kov\'acs, Peter Lepage, and
D.K.~Sinclair for a number of illuminating conversations on the topics
discussed in this paper.  I would also like to thank Andreas Kronfeld
and Yigal Shamir for their comments on earlier versions of this paper. 
Special thanks are due to Maarten Golterman and Eve Kov\'acs for their
extensive and detailed comments on several versions of the manuscript.
This work was supported in part by the U.S. Department of Energy,
Division of High Energy Physics, under Contract W-31-109-ENG-38.

%%%%%%%%%%%%%%%%%%%% appendices %%%%%%%%%%%%%%%%%%%%%%%%%%%%%%%%%%%%%%%%%%

\appendix
\section{COMPUTATION OF THE ANOMALY}
\label{app:anomaly}

In this Appendix we present a calculation of the gauge variation of the
odd-parity parts of fermion loops in the presence of a background gauge
field in four dimensions \cite{prev-anom}.  For simplicity, we
restrict ourselves to the case in which the Wilson term has not been
gauged.  If one includes the gauging of the Wilson term
(\ref{wilson-gauge}), then one must consider additional contributions to
the gauge variation involving $\Lambda$-gauge-field-fermion vertices. 

We will use repeatedly the fact that a trace containing an odd number of
$\gamma_5$'s is nonvanishing only if it contains four factors that are
linearly independent combinations of the the matrices $\gamma_1$,
$\gamma_2$, $\gamma_3$, $\gamma_4$.  These linearly independent
combinations can come from three sources:  the $\gamma$~matrices
associated with naive vertices in the loop, the $\gamma$~matrices
associated with external momenta in propagators, and the
$\gamma$~matrices associated with the loop momentum in propagators. 

In order to expose the external momenta, we expand the propagators and
vertices in a Taylor series in the external momenta times the lattice
spacing $a$.  We can use the result, derived in
Section~\ref{sec:divergent}, that a loop containing a $\Lambda$~vertex
receives a nonvanishing contribution in the limit $a\rightarrow 0$ only
from the region of integration in which the magnitude of the loop
momentum is of order $\pi/a$. In this region, it is easy to see, from
the discussion in Section~\ref{sec:divergent} and the fact that the
external momenta are assumed to be much smaller in magnitude than the
cutoff $\pi/a$, that the $n$th term in the Taylor expansion has a
relative suppression factor $a^n$. Thus, for a loop with degree of
divergence $D$, terms in the Taylor expansion containing more than $D$
factors of the external momenta do not receive a nonvanishing
contribution from the region of large loop momentum in the limit
$a\rightarrow 0$. Therefore, we retain only the first $D$ terms in the
Taylor expansion. For these terms, it can be seen, from the discussion
in Section~\ref{sec:divergent}, that the region of integration in which
the magnitude of the loop momentum is much less than $\pi/a$ gives a
negligible contribution.  Thus, we can extend the range of the
integration to the entire Brillouin zone. 

We also note that the $\gamma$~matrices associated with the loop
momentum can never contribute the required linearly independent factors:
if a term contains an odd number of $\gamma$-matrix factors associated
with the loop momentum, it gives a vanishing contribution because the
integrand is an odd function of the loop momentum; if a term contains an
even number of $\gamma$-matrix factors associated with the external
momentum, these factors can be brought together by using the
anti-commutation relations and eliminated by using $(\gamma\cdot
a)^2=a^2$. 

Armed with these facts, let us consider in turn the various
contributions up to those containing four external gauge fields. 

The contribution involving one $\Lambda$~vertex and no external gauge
fields vanishes by Abelian charge-conjugation symmetry. 

Next consider the contribution involving one $\Lambda$~vertex and one 
external gauge field.  If the gauge-field vertex is a naive vertex, it 
can contribute one of the linearly independent $\gamma$-matrix factors.  
The one independent external momentum can contribute another.  However, 
that is not enough to saturate a trace containing an odd number of 
$\gamma_5$'s.

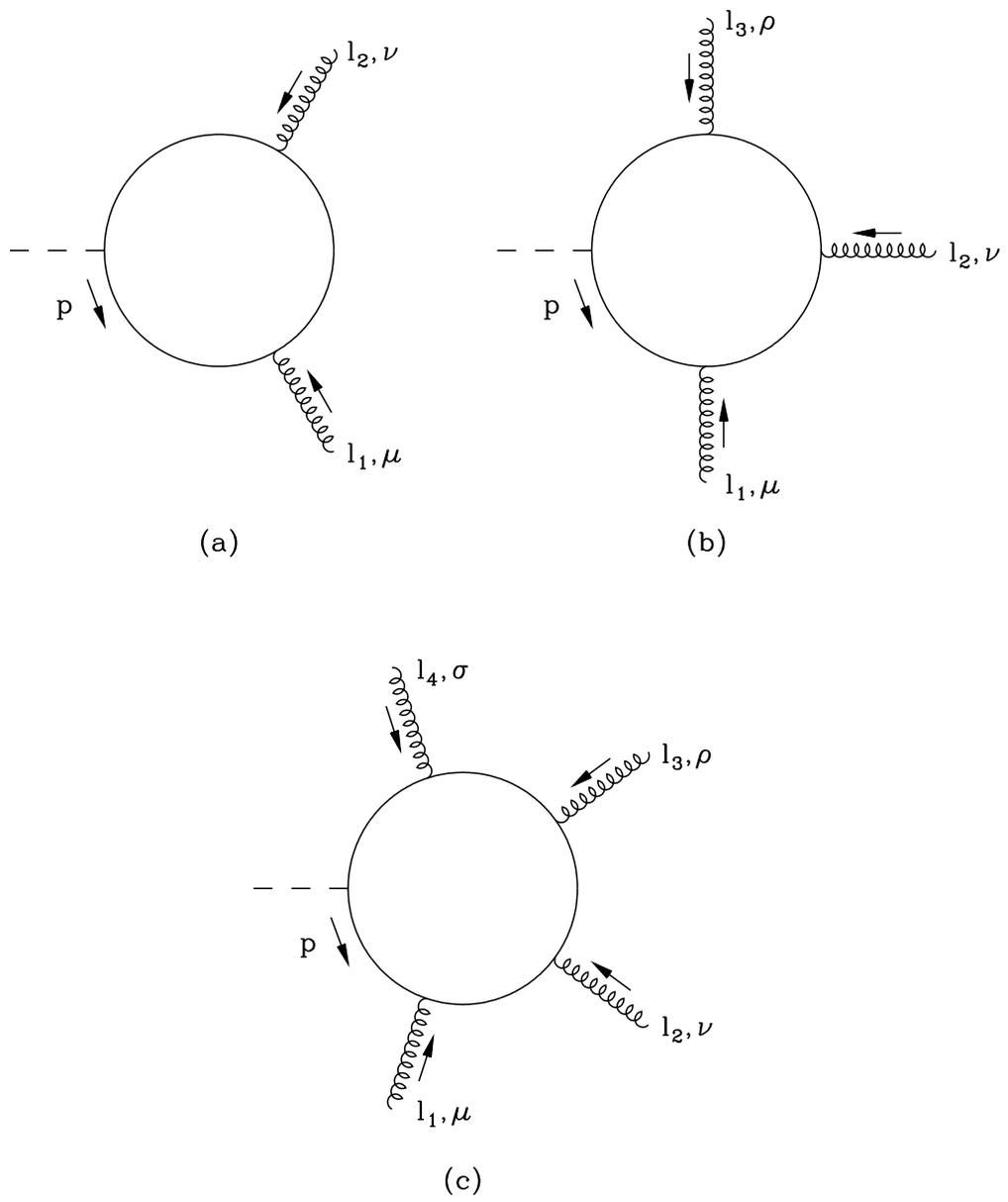
\begin{figure}
\caption{Diagrams that contribute to the ABJ anomaly in four
dimensions.} 
\label{fig:anomaly}
\end{figure}

In the contribution involving one $\Lambda$~vertex and two external
gauge fields, we can have at most two factors of external momentum in
the Taylor expansion and still obtain a nonvanishing contribution in the
limit $a \rightarrow 0$. Then, in order to obtain a nonvanishing trace,
we must take all of the gauge-field-fermion vertices to be of the type
${\cal V}^{(1)}$, which involves a single gauge field, and we must
retain terms proportional to the external momentum only in the Taylor
expansions of the propagators. The nonvanishing contribution then comes
from the diagram of Fig.~\ref{fig:anomaly}(a), whose amplitude we denote
by $A^{(2)}$, plus the diagrams obtained by permuting the gauge fields. 
That contribution is given by 
\begin{mathletters}
\label{anomaly}
\begin{eqnarray}
&&\lim_{a\rightarrow 0}
[A^{(2)}_{\mu\nu}(l_1,\mu,b;l_2,\nu,c)
+\hbox{perm}(l_1,\mu,b;l_2,\nu,c)]\nonumber\\
&&=g^2\int_{-\pi}^\pi{d^4 p\over (2\pi)^4}\,{\rm Tr}\,\Biggl\{iT_a\gamma_5 
M(p)\left[{\partial\over \partial (ap_\rho)}S_F^W(p)\right]
(l_{1}+l_{2})_\rho a V_\nu^{(1)N}(p)\nonumber\\
&&\qquad\times 
\left[{\partial\over \partial (ap_\sigma)}S_F^W(p)\right]l_{1\sigma}a
V_\mu^{(1)N}(p)S_F^W(p)\Biggr\}_{\rm odd}
+\hbox{perm}(l_1,\mu,b;l_2,\nu,c)\nonumber\\
&&=g^2I_{\mu\nu\rho\sigma}^{(2)}l_{2\rho} l_{1\sigma}
(1/2){\rm Tr}\,(T_a\{T_b,T_c\})
+\hbox{perm}(l_1,\mu,b;l_2,\nu,c).
\label{anomaly2}
\end{eqnarray}
Here, sums over repeated indices are understood. The subscript ``odd''
on the trace means that we retain only those terms that contain an odd
number of $\gamma_5$'s, and ``perm'' means permutations of the symbols
separated by semicolons, {\it i.e.,} permutations of the gauge fields.
In the last line we have used the fact, which follows from the
computation of the trace, that $I_{\mu\nu\rho\sigma}^{(2)}$ is
proportional to $\epsilon_{\mu\nu\rho\sigma}$. 

A similar analysis shows that the nonvanishing contribution involving
one $\Lambda$~vertex and three external gauge fields is given in the
limit $a\rightarrow 0$ by the diagram of Fig.~\ref{fig:anomaly}(b),
whose amplitude we denote by $A^{(3)}$, plus the diagrams obtained by
permuting the gauge fields. In this case, $D=1$, and so we retain only
one power of the external momentum in the Taylor expansion. The result
is 
\begin{eqnarray}
&&\lim_{a\rightarrow 0}
[A^{(3)}_{\mu\nu\rho}(l_1,\mu,b;l_2,\nu,c;l_3,\rho,d)
+\hbox{perm}(l_1,\mu,b;l_2,\nu,c;l_3,\rho,d)]
\nonumber\\
&&=g^3\int_{-\pi}^\pi{d^4 p\over (2\pi)^4}\,{\rm Tr}\,\Biggl\{
iT_a\gamma_5 M(p)
\left[{\partial\over \partial (ap_\sigma)}S_F^W(p)\right]
(l_{1}+l_{2}+l_{3})_\sigma a\nonumber\\
&&\hskip 2.5cm\qquad \times V_\rho^{(1)N}(p)S_F^W(p)
V_\nu^{(1)N}(p)S_F^W(p)
V_\mu^{(1)N}(p)S_F^W(p)\nonumber\\
&&\hskip 3.0cm +iT_a\gamma_5 M(p)S_F^W(p) V_\rho^{(1)N}(p)
\left[{\partial\over \partial (ap_\sigma)}S_F^W(p)\right]
(l_{1}+l_{2})_\sigma a\nonumber\\
&&\hskip 3.0cm\qquad \times V_\nu^{(1)N}(p)S_F^W(p)
V_\mu^{(1)N}(p)S_F^W(p)\nonumber\\
&&\hskip 3.0cm +iT_a\gamma_5 M(p)S_F^W(p) V_\rho^{(1)N}(p)
S_F^W(p)V_\nu^{(1)N}(p)\nonumber\\
&&\hskip 3.0cm\qquad \times
\left[{\partial\over \partial (ap_\sigma)}S_F^W(p)\right]l_{1\sigma}a
V_\mu^{(1)N}(p)
S_F^W(p)\Biggr\}_{\rm odd}{\rm Tr}\,(T_aT_bT_cT_d)\nonumber\\
&&\hskip 3.0cm
+\hbox{perm}(l_1,\mu,b;l_2,\nu,c;l_3,\rho,d)\nonumber\\
&&=g^3I_{\mu\nu\rho\sigma}^{(3)}
(l_1+l_2+l_3)_\sigma{\rm Tr}\,(T_aT_bT_cT_d)
+\hbox{perm}(l_1,\mu,b;l_2,\nu,c;l_3,\rho,d)
\nonumber\\
&&=g^3I_{\mu\nu\rho\sigma}^{(3)}
\Bigl[l_{1\sigma}(1/4){\rm Tr}\,(T_a\{T_b,[T_c,T_d]\})
+l_{3\sigma}(1/4){\rm Tr}\,(T_a\{T_d,[T_b,T_c]\})\nonumber\\
&&\qquad 
+l_{2\sigma}(1/2){\rm Tr}\,(T_a\{T_b,[T_c,T_d]\}+T_a\{T_d,[T_b,T_c]\}
+T_a\{T_c,[T_b,T_d]\})\Bigr]\nonumber\\
&&\qquad +\hbox{perm}(l_1,\mu,b;l_2,\nu,c;l_3,\rho,d).
\label{anomaly3}
\end{eqnarray}
Here, we have used the facts that only the first Dirac trace is nonzero and 
that it is proportional to $\epsilon_{\mu\nu\rho\sigma}$.

It is easily seen that the contribution involving one $\Lambda$~vertex
and four external gauge fields is given in the limit $a\rightarrow 0$ by
the diagram of Fig.~\ref{fig:anomaly}(c), whose amplitude we denote by
$A^{(4)}$, plus the diagrams obtained by permuting the gauge fields. In
this case, $D=0$, and so we set the external momenta equal to zero.  The
result is 
\begin{eqnarray}
&&\lim_{a\rightarrow 0}[A^{(4)}_{\mu\nu\rho\sigma}
(l_1,\mu,b;l_2,\nu,c;l_3,\rho,d;l_4,\sigma,e)
+\hbox{perm}(l_1,\mu,b;l_2,\nu,c;l_3,\rho,d;l_4,\sigma,e)]\nonumber\\
&&=g^4\int_{-\pi}^\pi{d^4 p\over (2\pi)^4}\,{\rm Tr}\,[
iT_a\gamma_5 M(p)S_F^W(p)V_\sigma^{(1)N}(p)S_F^W(p)
V_\rho^{(1)N}(p)S_F^W(p)V_\nu^{(1)N}(p)S_F^W(p)
\nonumber\\
&&\hskip 3.0cm\qquad \times 
V_\mu^{(1)N}(p)S_F^W(p)]_{\rm odd}
{\rm Tr}\,(T_aT_bT_cT_dT_e)
\nonumber\\
&&\hskip 3.0cm
+\hbox{perm}(l_1,\mu,b;l_2,\nu,c;l_3,\rho,d;
l_4,\sigma,e)\nonumber\\
&&=[g^4I_{\mu\nu\rho\sigma}^{(4)}
(1/8){\rm Tr}\,(T_a\{[T_b,T_c],[T_d,T_e]\})\nonumber\\
&&\qquad+\hbox{perm}(l_1,\mu,b;l_2,\nu,c;l_3,\rho,d;
l_4,\sigma,e)].
\label{anomaly4}
\end{eqnarray}
\end{mathletters}%
Again we have used the fact that the Dirac trace is proportional to 
$\epsilon_{\mu\nu\rho\sigma}$.  In fact, direct computation of the trace 
shows that 
\begin{equation}
I^{(4)}_{\mu\nu\rho\sigma}=0.
\end{equation}
We see that the odd-parity contributions from the fermion loops all vanish
in the limit $a\rightarrow 0$ if the anomaly-cancellation condition
(\ref{anom-cancel}) is satisfied. 

Now let us sketch a method by which the calculation of $I^{(2)}$ and
$I^{(3)}$ can be completed. If we drop the color factors in
(\ref{anomaly}), then the resulting expressions correspond to the
calculation of the gauge variations in an Abelian theory.  Since
$I^{(2)}$ and $I^{(3)}$ are symmetric under cyclic permutations of the
gauge fields, we can compute them by considering cyclic permutations of
the Abelian expressions for the gauge variations. 

Consider the quantity $\tilde\Gamma_{\alpha\mu\nu\ldots}^{(n)}$, which
is the Abelian amplitude associated with the odd-parity part of a
particular set of diagrams involving a fermion loop, $n$ gauge fields
(with indices $\alpha\mu\nu\ldots$), and no $\Lambda$~vertices. We
include in $\tilde\Gamma^{(n)}$ the diagram with no
multi-gauge-field-fermion vertices and the diagram with a single
two-gauge-field-fermion vertex involving the gauge fields with indices
$\alpha$ and $\mu$. We note the following relation between the Abelian
gauge variation and $\tilde\Gamma^{(n)}$: 
\begin{eqnarray}
&&-(i/g)d_\alpha(k)[\tilde\Gamma_{\alpha\mu\nu\ldots}^{(n)}
(l_1,l_2,l_3,\ldots,l_{n-1})
+\hbox{cyclic perm}(l_1,\mu;l_2,\nu;\ldots)]\nonumber\\
&&=\tilde A_{\mu\nu\ldots}^{(n-1)}(l_1,l_2,\ldots,\l_{n-1})
+\hbox{cyclic perm}(l_1,\mu;l_2,\nu;\ldots),
\label{div-eq-gauge}
\end{eqnarray}
where the tildes denote the Abelian case, 
\begin{equation}
k=-\sum_{i=1}^{n-1}l_i, 
\end{equation}
and $d_\alpha$ is defined in (\ref{lat-mom}). This relation follows from
the fact that the left side of (\ref{div-eq-gauge}) is the gauge
variation that one obtains by taking 
\begin{equation}
A_\mu(x+a_\mu/2)\rightarrow A_\mu(x+a_\mu/2)
+(1/ag)[\Lambda(x)-\Lambda(x+a_\mu)],
\label{abelian-gauge-tx2}
\end{equation}
which is equivalent to (\ref{u-gauge-tx}) in an Abelian theory, and
absorbing the transformation of the fermion fields (\ref{psi-gauge-tx})
into a change of variables in the path integral.  (One Fourier
transforms (\ref{abelian-gauge-tx2}) with respect to the coordinate of
the gauge field to obtain the left side of (\ref{div-eq-gauge}).) At a
graphical level, the relation (\ref{div-eq-gauge}) is obtained
by applying repeatedly the Feynman identity 
\begin{eqnarray}
d_\alpha(k)[V^{(1)N}(p,k)+V^{(1)W}(p,k)]&=&
-[iS_F^{W}(p+k)]^{-1}P_L+P_R[iS_F^{W}(p)]^{-1}\nonumber\\
&&\qquad -(1-P_L)M(p+k)+(1-P_R)M(p),
\label{feynman-ident}
\end{eqnarray}
as in textbook demonstrations of gauge invariance at the Feynman-graph
level.  The $M$ terms, of course, give the $\Lambda$~vertices on the
right side of (\ref{div-eq-gauge}). For the inverse propagator terms,
one does not find the simple pairwise cancellation that occurs in the
continuum theory because the lattice vertices are momentum dependent. It
follows from the recursion relation (\ref{recursion}) that this momentum
dependence is compensated by the contributions that one obtains by
contracting $d_\mu(k)$ with the two-gauge-field vertices. The result is
a complete cancellation of the inverse propagator terms.\footnote{If we
had gauged the Wilson term in the action, then there would be Wilson
vertices in the amplitudes, as well as naive vertices. The cancellation
of the inverse propagator terms would fail in the presence of the Wilson
vertices because they commute rather than anticommute with the
$\gamma_5$'s in the inverse-propagator terms in (\ref{feynman-ident}).
Consequently, a more complicated identity than (\ref{div-eq-gauge}),
involving $\Lambda$-gauge-field-fermion vertices, would be obtained.} 

Now, $\tilde\Gamma^{(n)}$ receives no contributions from the region of
integration in which the magnitude of the loop momentum is of order
$\pi/a$.  This follows from the fact, discussed in
Section~\ref{sec:counting-a-f}, that the odd-parity parts of loops have
no renormalization counterterms that are invariant under cyclic
permutations of the gauge fields. It can also be seen by expanding
$\tilde\Gamma^{(n)}$ in a Taylor series in the external momenta. The
first $5-n$ terms in the expansion have a vanishing trace under cyclic
permutations of the gauge fields; the remainder in the expansion is
suppressed by powers of $a$ when the magnitude of the loop momentum is
of order $\pi/a$. We conclude that we can evaluate $\tilde\Gamma^{(n)}$
(including all permutations of the gauge fields) by taking the limit
$a\rightarrow 0$ in the propagators and vertices.  The result is just
the continuum expression. Thus, 
\begin{eqnarray}
&&\lim_{a\rightarrow 0}
\tilde A_{\mu\nu\ldots}^{(n-1)}(l_1,l_2,\ldots,\l_{n-1})
+\hbox{cyclic perm}(l_1,\mu;l_2,\nu;\ldots)\nonumber\\
&&=-(i/g)k_\alpha[\tilde\Gamma_{\alpha\mu\nu\ldots}^{(n){\rm cont}}
(l_1,l_2,l_3,\ldots,l_{n-1})
+\hbox{cyclic perm}(l_1,\mu;l_2,\nu;\ldots)].
\label{div-eq-cont}
\end{eqnarray}
The right side of (\ref{div-eq-cont}) is just the continuum expression
for the ABJ anomaly.  We can evaluate it by considering the gauge
variation of the continuum action in the presence of a UV regulator.  If
we impose a Pauli-Villars regulator, then we obtain expressions that are
identical to those in (\ref{anomaly2}) and (\ref{anomaly3}), except that
there are no color factors, the Wilson mass $M(p)$ is replaced
everywhere by the Pauli-Villars mass, the limit $a\rightarrow 0$ is
taken in the remaining terms in the propagators and vertices, and there
is a minus sign because one subtracts the massive
Pauli-Villars-regulator contribution. The results are 
\begin{mathletters}
\label{anomaly-int}
\begin{eqnarray}
I_{\mu\nu\rho\sigma}^{(2)}&=&-i/(24\pi^2)\epsilon_{\mu\nu\rho\sigma},\\
I_{\mu\nu\rho\sigma}^{(3)}&=&i/(48\pi^2)\epsilon_{\mu\nu\rho\sigma},
\end{eqnarray}
\end{mathletters}%
which, upon continuation to Minkowski space, can be seen to be in
agreement with previous calculations of the gauge (consistent) anomaly
\cite{bardeen}. 

This result is actually independent of the choice of UV regulator. As we
have already mentioned, if one assumes symmetry under cyclic
permutations of the gauge fields, then there are no renormalization
counterterms for the odd-parity parts of the ordinary fermion-loop
amplitudes (those associated with diagrams that do not contain
$\Lambda$~vertices).  The absence of counterterms guarantees that the
amplitudes themselves are regulator independent. Furthermore, the
anomaly can be obtained from the amplitudes by varying the gauge fields
according to (\ref{u-gauge-tx}) and absorbing the transformation of the
fermion fields (\ref{psi-gauge-tx}) into a change of variables in the
path integral, as was discussed explicitly for the Abelian case in
reference to (\ref{div-eq-gauge}). Therefore, the anomaly is also
regulator independent. In particular, we would have obtained the result
(\ref{anomaly-int}) had we chosen to retain the gauging of the Wilson
term in the action.

%%%%%%%%%%%%%%%%%%%%%%%%%%%%%%  REFERENCES  %%%%%%%%%%%%%%%%%%%%%%%%%%%%%%

%%%%%%%%%%%%%%%%%%%%%%%%%%%%  FIGURE CAPTIONS  %%%%%%%%%%%%%%%%%%%%%%%%%%%%%%

\begin{references}

\bibitem{shamir}
Y.~Shamir, in {\it Lattice~'95}, Proceedings of the International 
Symposium, Melbourne, Australia, edited by T.D.~Kieu, {\it et al.} 
[Nucl.\ Phys.\ B (Proc.\ Suppl.) {\bf 47}, 212 (1996)],
hep-lat/9509023.

\bibitem{slavnov}
A.A.~Slavnov, Phys.\ Lett.\ B {\bf 319}, 231 (1993); {\bf 348}, 553 
(1995).

\bibitem{zenkin}
S.~Zenkin, Phys.\ Lett.\ B {\bf 366}, 261 (1996).

\bibitem{bietenholz-weise}
W.~Bietenholz and U.-J.~Wiese, MIT preprint MIT-CTP-2423,
hep-lat/9503022.

\bibitem{eichten-preskill}
E.~Eichten and J.~Preskill, Nucl.\ Phys.\ {\bf B268}, 179 (1986).

\bibitem{smit-swift}
P.~Swift, Phys.\ Lett.\ {\bf 145B}, 256 (1984); J.~Smit, Acta.\ Phys.\ 
Pol.\ B {\bf 17}, 531 (1986).

\bibitem{smit}
J.~Smit,
in {\it Field Theory on the Lattice}, Proceedings of the International
Symposium, Seillac, France, edited by A.~Billoire {\it et al.} [Nucl.\
Phys.\ B (Proc.\ Suppl.) {\bf 4}, 451 (1988)]. 

\bibitem{staggered-fermion}
W.~Bock, J.~Smit, and J.C.~Vink, Nucl.\ Phys.\ {\bf B414}, 73 (1994);
{\bf B416}, 645 (1994). 

\bibitem{eichten-preskill-n}
M.F.L.~Golterman and D.N.~Petcher, in {\it Lattice~'91}, Proceedings of
the International Symposium, Tsukuba, Japan, edited by M.~Fukugita {\it
et al.} [Nucl.\ Phys.\ B (Proc.\ Suppl.) {\bf 26}, 486 (1992)];
M.F.L.~Golterman, D.N.~Petcher, and E.~Rivas, in {\it Nonperturbative
Aspects of Chiral Gauge Theories}, Proceedings of the Topical Workshop,
Rome, Italy, 1992, edited by L.~Maiani {\it et al.} [Nucl.\ Phys.\ B
(Proc.\ Suppl.) {\bf 29}, 193 (1992)]; M.F.L.~Golterman, D.N.~Petcher,
and E.~Rivas, Nucl.\ Phys.\ {\bf B395}, 596 (1993). 

\bibitem{smit-swift-n}
M.F.L.~Golterman, D.N.~Petcher, and J.~Smit, Nucl.\ Phys.\ {\bf B370},
51 (1992); W.~Bock, A.K.~De, and J.~Smit, Nucl.\ Phys.\ {\bf B388}, 243
(1992); M.F.L.~Golterman and D.N.~Petcher, in {\it Lattice~'91},
Proceedings of the International Symposium, Tsukuba, Japan, edited by
M.~Fukugita {\it et al.} [Nucl.\ Phys.\ B (Proc.\ Suppl.) {\bf 26}, 483
(1992)]. 

\bibitem{domain-wall}
D.B.~Kaplan, Phys.\ Lett.\ B {\bf 288}, 342 (1992); in {\it Lattice~'92},
Proceedings of the International Symposium, Amsterdam, The Netherlands,
edited by J.~Smit and P.~van Baal [Nucl.\ Phys.\ B (Proc.\ Suppl.) {\bf
30}, 597 (1993)]. 

\bibitem{narayanan-neuberger}
R.~Narayanan and H.~Neuberger, Phys.\ Lett.\ B {\bf 302}, 62 (1993);
Nucl.\ Phys.\ {\bf B412}, 574 (1994); 
{\bf B443}, 305 (1995).

\bibitem{domain-wall-p}
K.~Jansen, Phys.\ Lett.\ B {\bf 288}, 348 (1992);
M.F.L.~Golterman, and K.~Jansen, and D.B.~Kaplan, Phys.\ Lett.\ B {\bf 
301}, 219 (1993); 
Y.~Shamir, Nucl.\ Phys.\ {\bf B406}, 90 (1993); 
{\bf B417}, 167 (1994); 
S.~Chandrasekharan, Phys.\ Rev.\ D {\bf 49}, 1980 (1994);
S.~Aoki and H.~Hirose, Phys.\ Rev.\ D {\bf 49}, 2604 ((1994);
M.~Creutz and I.~Horvath, in {\it Lattice~'93}, Proceedings of the
International Symposium, Dallas, Texas, edited by T.~Draper {\it et al.}
[Nucl.\ Phys.\ B (Proc.\ Suppl.) {\bf 34}, 799 (1994)]; 
Phys.\ Rev.\ D {\bf 50}, 2297 (1994); 
V.~Furman and Y.~Shamir, Nucl.\ Phys.\ {\bf B439}, 54 (1995); 

\bibitem{overlap-p}
R.~Narayanan and H.~Neuberger, Phys.\ Lett.\ B {\bf 353}, 507 (1993);
{\bf 348}, 549 (1995); 
Phys.\ Rev.\ Lett.\ {\bf 71}, 3251 (1993); 
in {\it Lattice~'93}, Proceedings of the International Symposium,
Dallas, Texas, edited by T.~Draper {\it et al.} [Nucl.\ Phys.\ B (Proc.\
Suppl.) {\bf 34}, 587 (1994)]; 
S.~Aoki and R.B.~Levien, Phys.\ Rev.\ D {\bf 51}, 3790 (1995); 
S.~Randjbar-Daemi and J.~Strathdee, Nucl.\ Phys.\ {\bf B443}, 386
(1995); Phys.\ Rev.\ D {\bf 51}, 6617 (1995); Phys.\ Lett.\ B {\bf 348},
543 (1995); R.~Narayanan, H.~Neuberger, and P.~Vranas, Phys.\ Lett.\
B {\bf 353}, 507 (1995). 
T.~Kawano and Y.~Kikukawa, Kyoto University preprint KUNS-1317,
hep-lat/9501032. 

\bibitem{domain-wall-n}
M.F.L.~Golterman, K.~Jansen, D.N.~Petcher, and J.C.~Vink, Phys.\ Rev.\ D 
{\bf 49}, 1606, (1994); M.F.L.~Golterman and Y.~Shamir, Phys.\ Rev.\ D 
{\bf 51}, 3026 (1995).

\bibitem {overlap-n}
Y.~Shamir, Nucl.\ Phys.\ {\bf B417}, 167 (1994);
M.F.L.~Golterman and Y.~Shamir, Phys.\ Lett.\ B {\bf 353}, 84 (1995);
{\bf B359}, 422(E) (1995); 
in {\it Lattice~'95}, Proceedings of the International 
Symposium, Melbourne, Australia, edited by T.D.~Kieu, {\it et al.} 
[Nuclear Physics B (Proc.\ Suppl.) {\bf 47}, 603 (1996)],
hep-lat/9509027.

\bibitem{wilson}
K.G.~Wilson, Phys.\ Rev.\ D {\bf 10}, 2445 (1974); in {\it New Phenomena 
in Subnuclear Physics}, edited by A.~Zichichi (Plenum, New York, 1977).

\bibitem{lat90}
G.T.~Bodwin and E.V.~Kov\'acs, in {\it Lattice '90}, Proceedings of the 
International Symposium, Tallahassee, Florida, edited by U.M.~Heller, 
A.D.~Kennedy, and S.~Sanielevici [Nucl.\ Phys.\ B (Proc.\ Suppl.) {\bf 
20}, 546 (1991)].

\bibitem{dpf91}
G.T.~Bodwin and E.V.~Kov\'acs, in {\it The Vancouver Meeting---Particles 
and Fields '91}, Proceedings of the Conference, Vancouver, Canada, 
edited by D.~Axen, D.~Bryman, and M.~Comyn (World Scientific, Singapore, 
1992).

\bibitem{lat92}
G.T.~Bodwin and E.V.~Kov\'acs, 
in {\it Lattice '92}, Proceedings of the International Symposium,
Amsterdam, The Netherlands, edited by J.~Smit and P.~van Baal [Nucl.\
Phys.\ B (Proc.\ Suppl.) {\bf 30}, 617 (1993)]. A method that achieves an 
equivalent result by making use of auxiliary fermion species is 
presented in 
Refs.~\cite{lat90}~and~\cite{dpf91}.

\bibitem{rome}
A.~Borrelli, L.~Maiani, G.C.~Rossi, R.~Sisto, and M.~Testa,
Nucl.\ Phys.\ {\bf B333}, 335 (1990).
L.~Maiani, G.C.~Rossi, and M.~Testa  Phys.\ Lett.\ B {\bf 261}, 479 (1991);
{\bf 292}, 397 (1992).

\bibitem{zaragoza}
J.L.~Alonso, Ph.~Boucaud, J.L.~Cort\'es, and E.~Rivas, 
in {\it Lattice~'89}, Proceedings of the International Symposium, Capri,
Italy, edited by A.~Petronzio {\it et al.} [Nucl.\ Phys.\ B (Proc.\
Suppl.) {\bf 17}, 461 (1990)]; Mod.\ Phys.\ Lett.\ A {\bf 5}, 275 (1990);
Phys.\ Rev.\ D {\bf 44}, 3258 (1991);
J.L.~Alonso, Ph.~Boucaud, F.~Lesmes, and A.J.~van der Sijs, 
Nucl.\ Phys.\ {\bf B457}, 175 (1995).

\bibitem{alvarez-gaume-pietra}
L.~\'Alvarez-Gaum\'e and S.~Della Pietra, 
in {\it Recent Developments in Quantum Field Theory}, edited by
J.~Ambjorn, B.J.~Durhuus, and J.L.~Petersen (North Holland, Amsterdam,
1985). 

\bibitem{gockeler-schierholz}
M.~G\"ockeler and G.~Schierholz,
in {\it Lattice '92}, Proceedings of the International Symposium,
Amsterdam, The Netherlands, edited by J.~Smit and P.~van Baal [Nucl.\
Phys.\ B (Proc.\ Suppl.) {\bf 30}, 609 (1993)]. 

\bibitem{effective-actions}
A number of studies over the years have clarified the nature of the 
effective action generated by chiral fermions.  See, for example, 
S.~Aoki, Phys.\ Rev.\ Lett.\ {\bf 60}, 2109 (1988);
Phys.\ Rev.\ D {\bf 38}, 618 (1988);
{\bf 42}, 2806 (1990);
K.~Funakubo and T.~Kashiwa, Phys.\ Rev.\ Lett.\ {\bf 60}, 2113 (1988).

\bibitem{anlreport}
G.~Bodwin, 
in ``High Energy Physics Division Semiannual Report of Research
Activities, January 1, 1993--June 30, 1993,'' Argonne National Laboratory
Report ANL-HEP-TR-93-88, 1993, p.~43.

\bibitem{thooft}
G.~'t~Hooft,
Phys.\ Lett.\ B {\bf 349}, 491 (1995).

\bibitem{hsu}
S.D.H.~Hsu,
Yale preprint YCTP-P5-95, hep-th/9503058.

\bibitem{kronfeld}
A.S.~Kronfeld,
Fermilab preprint FERMILAB-PUB-95/073-T, hep-lat/9504007.

\bibitem{hernandez-sundrum}
Nucl.\ Phys.\ {\bf B455}, 287 (1995).

\bibitem{karsten-smit}
L.H.\ Karsten and J.~Smit,
Nucl.\ Phys.\ {\bf B183}, 103 (1981).

\bibitem{no-go}
H.B.~Nielsen and M.~Ninomiya,
Nucl.\ Phys.\ {\bf B185}, 20 (1981); {\bf B193}, 173 (1981);
Phys.\ Lett.\ {\bf 105B}, 219 (1981);
M.~Ninomiya and C.-I.~Tan,
Phys.\ Rev.\ Lett.\ {\bf 53}, 1611 (1984).

\bibitem{adler-bardeen}
S.L.~Adler and W.A.~Bardeen,
Phys.\ Rev.\ {\bf 184}, 1848 (1969).

\bibitem{bodwin-kovacs-schwinger}
G.T.~Bodwin and E.V.~Kov\'acs,
Phys.\ Rev.\ D {\bf 35}, 3198 (1987).

\bibitem{witten}
E.~Witten,
Phys.\ Lett.\ {\bf 117B}, 324 (1982).

\bibitem{adler-bardeen2}
S.L.~Adler and W.A~.Bardeen,
Phys.\ Rev.\ {\bf 182}, 1517 (1969).

\bibitem{gockeler-et-al}
M.~G\"ockeler, A.S.~Kronfeld, G.~Schierholz, and 
U.-J.~Wiese, Nucl.\ Phys.\ {\bf B404}, 839 (1993).

\bibitem{golterman-petcher}
M.F.L.~Golterman and D.~N.~Petcher,
Phys.\ Lett.\ B {\bf 225}, 159 (1989).

\bibitem{wetzel}
W.~Wetzel, 
Phys.\ Lett.\ {\bf 136B}, 407 (1984).

\bibitem{eguchi-kawamoto}
T.~Eguchi and N.~Kawamoto,
Nucl.\ Phys.\ {\bf B237}, 609 (1984).

\bibitem{dugan-manohar}
A discussion of the baryon-number current on the lattice and its
renormalization counterterms has been given by M.J.~Dugan and
A.V.~Manohar, Phys.\ Lett.\ B {\bf 265}, 137 (1991). 

\bibitem{thooft-instantons}
G.~'t~Hooft,
Phys.\ Rev.\ D {\bf 14}, 3432 (1976); {\bf 18}, 2199 (1978).

\bibitem{montvay}
A two-dimensional chiral Schwinger model with lattice gauge fields and
continuum-regulated fermion fields has been considered by I.~Montvay
in CERN preprint CERN-TH/95-123, hep-lat/9505015.

\bibitem{prev-anom}
A calculation of the anomaly for Wilson fermions has been given
previously in Ref.~\cite{karsten-smit}; in A.~Coste,
C.~Korthals-Altes, and O.~Napoly, Nucl.\ Phys.\ {\bf B289}, 645 (1987); 
and in S.~Aoki, Phys.\ Rev.\ D {\bf 35}, 1435 (1987).

\bibitem{bardeen}
W.A.~Bardeen, Phys.\ Rev.\ {\bf 184}, 1848 (1969).

\end{references}
\end{document}